\documentclass[]{imsart}

\usepackage[T1]{fontenc}
\usepackage{amsthm}
\usepackage{bm}
\usepackage{amsmath}
\usepackage{amssymb}
\usepackage{natbib}
\usepackage{mathtools}
\usepackage{graphicx}
\usepackage{subcaption}
\usepackage{xcolor}
\usepackage{booktabs}
\usepackage{xr-hyper}
\usepackage[margin=1in]{geometry}
\usepackage[colorlinks,citecolor=blue,urlcolor=blue,filecolor=blue,backref=page]{hyperref}

\makeatletter
\newcommand*{\addFileDependency}[1]{
  \typeout{(#1)}
  \@addtofilelist{#1}
  \IfFileExists{#1}{}{\typeout{No file #1.}}
}
\makeatother

\newcommand{\Cov}{\mathrm{Cov}}
\newcommand{\Var}{\mathrm{Var}}
\newcommand{\StdDev}{\mathrm{StdDev}}
\DeclarePairedDelimiter\abs{\lvert}{\rvert}%

\startlocaldefs
\numberwithin{equation}{section}
\theoremstyle{plain}

\endlocaldefs

\begin{document}

\begin{frontmatter}

\title{Fully Bayesian inference for spatiotemporal data with the multi-resolution approximation}
\runtitle{Fully Bayesian inference with the MRA}

\begin{aug}

\author{\fnms{Luc} \snm{Villandr\'{e}}\thanksref{addr1,t1}\ead[label=e1]{luc.villandre@hec.ca}},
\author{\fnms{Jean-François} \snm{Plante}\thanksref{addr1}\ead[label=e2]{jfplante@hec.ca}},
\author{\fnms{Thierry} \snm{Duchesne}\thanksref{addr2}\ead[label=e3]{thierry.duchesne@mat.ulaval.ca}}
\and
\author{\fnms{Patrick} \snm{Brown}\thanksref{addr3}
\ead[label=e4]{patrick.brown@utoronto.ca}}

\runauthor{Villandr\'{e} et al.}

\address[addr1]{Department of Decision Sciences, HEC Montréal, 3000 Chemin de la Côte-Sainte-Catherine, Montréal, QC, Canada, H3T 2A7
    \printead{e1} 
    \printead*{e2}
}

\address[addr2]{Department of Mathematics and Statistics, Université Laval, 1045, av. de la Médecine, Université Laval, Québec, QC, Canada, G1V 0A6
    \printead{e3}    
}

\address[addr3]{Department of Statistical Sciences, University of Toronto, Room 6018, Sidney Smith Hall
100 St. George St., Toronto, ON M5S 3G3
    \printead{e4}    
}

\thankstext{t1}{Corresponding author}

\end{aug}

\begin{abstract}
Large spatiotemporal datasets are a challenge for conventional Bayesian models because of the cubic computational complexity of the algorithms for obtaining the Cholesky decomposition of the covariance matrix in the multivariate normal density. Moreover, standard numerical algorithms for posterior estimation, such as Markov Chain Monte Carlo (MCMC), are intractable in this context, as they require thousands, if not millions, of costly likelihood evaluations. To overcome those limitations, we propose IS-MRA (\textit{Importance sampling - Multi-Resolution Approximation}), which takes advantage of the sparse inverse covariance structure produced by the Multi-Resolution Approximation (MRA) approach. IS-MRA is fully Bayesian and facilitates the approximation of the hyperparameter marginal posterior distributions. We apply IS-MRA to large MODIS Level 3 Land Surface Temperature (LST) datasets, sampled between May 18 and May 31, 2012 in the western part of the state of Maharashtra, India. We find that IS-MRA can produce realistic prediction surfaces over regions where concentrated missingness, caused by sizable cloud cover, is observed. Through a validation analysis and simulation study, we also find that predictions tend to be very accurate.   
\end{abstract}



\end{frontmatter}

\section{Introduction}

Remote sensing has resulted in spatiotemporal data being produced at a very quick pace. Gaussian Random Fields (GRF) are at the core of many models conventionally applied to such data, but tend to be cumbersome when the datasets are large. Indeed, likelihood evaluations involve the inverse of a covariance matrix whose size increases with the number of observations. The computational complexity of the algorithm for obtaining the Cholesky decomposition, involved in matrix inversion and computation of the determinant, is cubic in the number of rows or columns. It follows that it cannot handle large dense matrices. Algorithms that rely on a sizable number of likelihood evaluations, such as Markov Chain Monte Carlo (MCMC), also turn out to be impractical, as a single likelihood calculation can take up to several minutes.

This study proposes a new Bayesian method, called IS-MRA, that helps overcome the computational limitations of conventional approaches for modeling spatiotemporal data. It relies on the Multi-Resolution Approximation \citep{Katzfuss2017, Katzfuss2017b} (MRA), which it extends to hyperparameter posterior estimation, making it fully Bayesian. This extension represents the first main contribution of our work. It does so with the help of a novel importance sampling algorithm, whose formulation and implementation represent the second main contribution of our study. Finally, the original formulation of the MRA was restricted to spatial data. The extension to spatiotemporal data represents a practical contribution, as it requires considering additional hyperparameters, a non-trivial computational challenge in light of the limitations of numerical integration algorithms.

Scalable models in spatial statistics tend to rely on a combination of three strategies \citep{Heaton2019}. First, the \textit{low-rank approximation} strategy involves a dimension-reduction scheme for the covariance matrix, achieved through decomposition of the GRF into a finite sum of orthogonal terms whose coefficients are obtained by computing the values of basis functions. The predictive process \citep{Banerjee2008}, LatticeKrig \citep{Nychka2015}, and fixed-rank Kriging \citep{Cressie2008}, for example, use that approach. A second strategy involves imposing sparsity on the covariance or precision matrix, i.e. the inverse of the covariance matrix. This can be done by assuming conditional independence between some of the observations, such as in the spatial partitioning \citep{Heaton2017a} and Stochastic Partial Differential Equation (SPDE) approaches \citep{Lindgren2011}, or through \textit{tapering}, that is, multiplying the covariance function by a correlation function with compact support\citep{Furrer2006}. Finally, recent methodological development has focused on parallelisation. Meta-kriging \citep{Guhaniyogi2018} and the  parallel low-rank models by \cite{Katzfuss2017a} have used that approach . 

The MRA \citep{Katzfuss2017, Katzfuss2017b} mixes all three strategies. It is a special case of the general Vecchia approximation, which has been extended to non-Gaussian observations as well \citep{Katzfuss2017, Zilber2020}. The MRA has been applied to spatiotemporal data, albeit in the context of state-space models \citep{Jurek2018}. Unlike other low-rank approaches, the MRA does not produce overly smooth predictions. By imposing sparsity on the precision matrices, matrices of a much higher rank can be used without straining computational resources. The MRA is formulated in such a way that core computations in the likelihood and posterior evaluations can be performed in parallel, with little communication required between processes. Its performance in terms of predictive accuracy has also been shown to be excellent \citep{Heaton2019}. 

\subsection{MODIS land surface temperature data}

\begin{figure}[ht]
    \centering
    \includegraphics[width = 11cm]{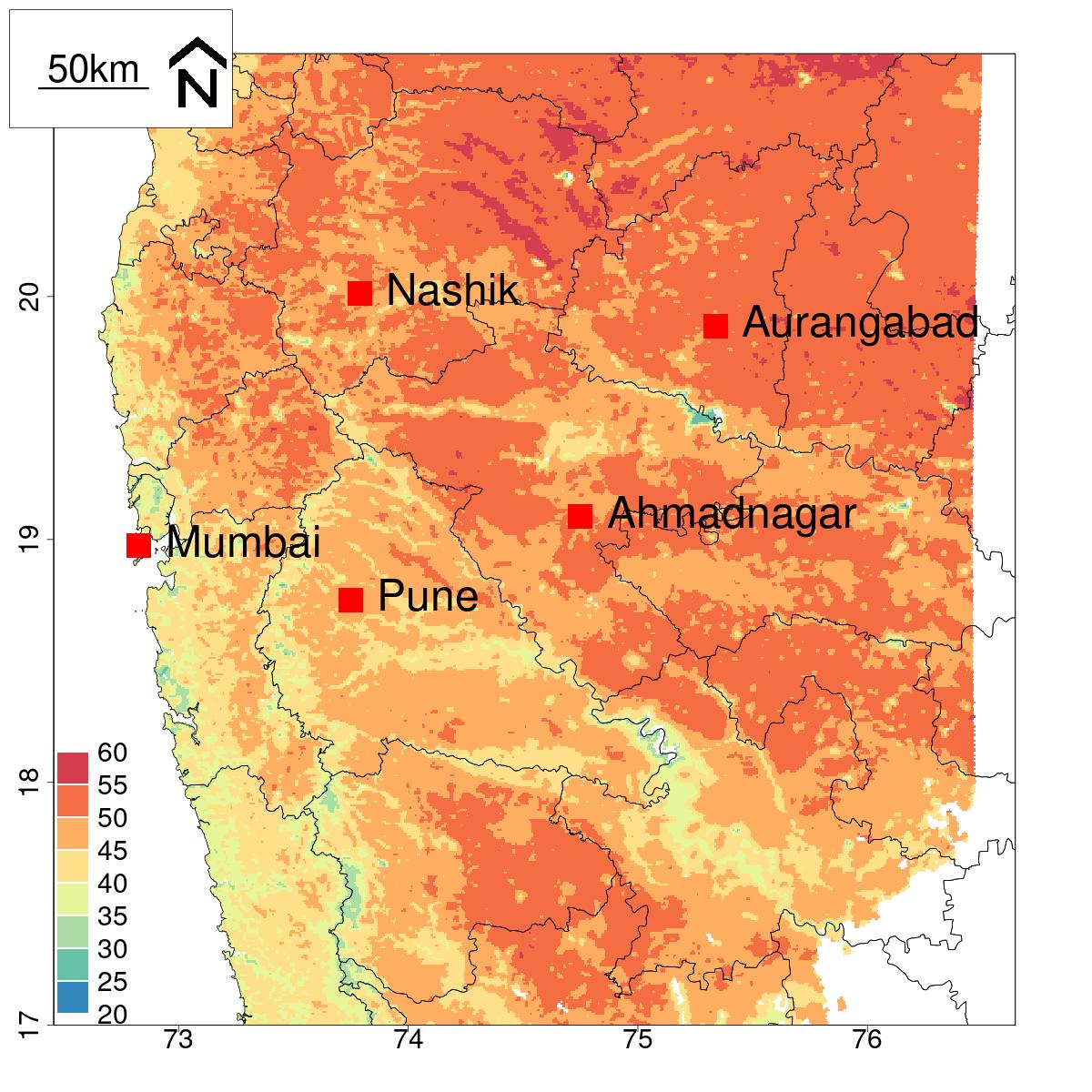}
    \caption{\textbf{Land surface temperatures (in degrees Celsius) recorded on May 29, 2012 over a 389 km $\times$ 445 km region in the western part of Maharashtra, India.} Also shown are district boundaries and markers for major cities.}
    \label{fig:temperaturesMay29}
\end{figure}

We will illustrate the use of IS-MRA on MODIS (Moderate Resolution Imaging Spectroradiometer) Level-3 land surface temperature (LST) data \citep{MODIS_Terra_LST, MODIS_Aqua_LST}. LST, also called land surface emissivity, is an indicator of ground temperature or brightness. LST does not correspond to \textit{air temperature}, measured by weather stations, but is highly correlated with it \citep{Mildrexler2011}. The MODIS imaging sensor has been loaded onto two satellites, Terra and Aqua, launched in 1999 and 2002, respectively. The raw spectroradiometry data collected by the spacecrafts is transformed into several gridded products, such as water vapor and aerosol levels, snow cover, and LST. The grid used for daily LST observations has a resolution of 1 km $\times$ 1 km. Missing values are however very common, as thick cloud formations may prevent the satellites from viewing the surface. 

This paper considers MODIS LST data collected on a plateau overlooking the Konkan coast in India, where Mumbai is located. LST data from May 29 2012, a sunny day with relatively few missing values, is shown in Fig. \ref{fig:temperaturesMay29}. The correspondence between geopolitical borders and temperature patterns is mostly due to the presence of hills or rivers, which serve as convenient boundaries for splitting a territory. A large patch of missing data appears in the southeastern corner. Inferring missing values is a problem that can be well addressed by Bayesian methods for spatiotemporal inference such as IS-MRA.

\section{Methods}

\subsection{Model}

Let the $i$'th observation, recorded at spatiotemporal coordinates $(\bm{s}_i, t_i)$, be denoted $Y_i$, the set of all sampled responses be denoted $\bm{y}$, and the corresponding matrix of covariates be denoted $\bm{X}$. We assume that
\begin{equation}
    Y_i \mid W(\bm{s}_i, t_i) \sim N[\bm{X}(\bm{s}_i, t_i) \bm{\beta} + W(\bm{s}_i, t_i),  \zeta^2], \label{eq:basicYi}
\end{equation}
with $W(\bm{s}, t)$ being a Gaussian Random Field (GRF) with mean $0$ and separable covariance function
\begin{equation}
 \Cov[W(\bm{s} + \bm{u}, t + v), W(\bm{s}, t)] = \sigma^2 \mbox{Mat\'{e}rn}(\abs{\bm{u}}; \rho, \nu) \exp(-\abs{v}/\phi), \label{eq:MaternCov} 
\end{equation}
$\phi$ being known as the \textit{temporal range} parameter. The expression $\mbox{Mat\'{e}rn}(\abs{\bm{u}}; \rho, \nu)$ denotes the Mat\'{e}rn covariance function,
\begin{equation}
 \mbox{Mat\'{e}rn}(\abs{\bm{u}}; \rho, \nu) = \frac{2^{1-\nu}}{\Gamma(\nu)}\Bigg(\sqrt{2\nu}\frac{\abs{\bm{u}}}{\rho}\Bigg)^\nu K_\nu\Bigg(\sqrt{2\nu}\frac{\abs{\bm{u}}}{\rho}\Bigg), \hspace{0.5cm} \rho > 0, \nu > 0,
\end{equation}
with $K_\nu(.)$ being the modified Bessel function of the second kind, $\abs{\bm{u}}$ being the norm of $\bm{u}$, and $\rho$ and $\nu$ being known as the spatial range and differentiability parameters, respectively. The spatial coordinates are in longitude and latitude, and we use the Haversine distance (or great circle distance, see \cite{Sinnott1984}) formula to obtain $\abs{\bm{u}}$. We fix $\nu$ at $1.5$, following the advice of \citet{Stein2012}. Fixed-effects coefficients $\bm{\beta}$ have a multivariate Gaussian prior with mean $\bm{0}$ and variance $\Sigma_{\bm{\beta}} = 100 \bm{I}$. Variance parameters, expressed on the logarithmic scale, are denoted $\bar{\bm{\Psi}}$ $\equiv$ $\{\log(\sigma),$ $\log(\rho),$ $\log(\phi)\}$ and $\bm{\Psi} \equiv \{\bar{\bm{\Psi}}, \log(\zeta)\}$, and also have independent Gaussian priors. Although Eq. \ref{eq:MaternCov} denotes a separable covariance structure, any covariance function, including non-separable functions, can be used with this methodology.

\subsection{The spatiotemporal Multi-Resolution Approximation} \label{section:MRA}

Under the MRA \citep{Katzfuss2017}, the spatiotemporal domain is partitioned into $K$ levels of increasingly fine resolution. Write $R_0$ to refer to the entire domain, and $R_{1\ell}$, $\ell=1\ldots L_1$, to the level $1$ regions, which together form a partition of $R_0$. Each deeper level $k$ has regions $R_{km}$, $m = 1, \dots, L_k$, formed by splitting the regions defined at the previous resolution. The partitioning is therefore nested, with either $R_{km} \subset R_{k-1,l}$ or $R_{km} \cap R_{k-1,l} = \varnothing$. For notational simplicity, and unlike in \citet{Katzfuss2017}, we have dispensed with the explicit mention of the nesting through a third subscript. We present in Figure 1 in Appendix C an example of the nesting that should clarify the notation. We define sets of increasingly dense knot positions $\bm{\mathcal{Q}}_k, k = 1, \ldots, K$, as well as interpolants $\tau_k(\bm{s},t)$ and residuals $\delta_k(\bm{s},t)$, with 
\begin{align*}    
    \tau_0(\bm{s},t) &= E[W(\bm{s},t) \mid W(\bm{u},v) : \bm{u},v \in \bm{\mathcal{Q}}_0], \hspace{0.5cm} \\
    \delta_1(\bm{s},t) &= W(\bm{s},t) - \tau_0(\bm{s},t), \\
    \tau_k(\bm{s},t) &= E[\delta_k(\bm{s},t) \mid \delta_k(\bm{u},v) :  \bm{u},v \in \bm{\mathcal{Q}}_k \cap R_{k\ell}], \hspace{0.5cm}(\bm{s},t) \in R_{k\ell}, \\
    \delta_{k+1}(\bm{s},t) &= \delta_{k}(\bm{s},t) - \tau_k(\bm{s},t).
\end{align*} 
The multi-resolution approximation to $W(\bm{s}, t)$ is
\begin{equation*}
    \widetilde{W}(\bm{s},t) = \delta_K(\bm{s},t) + \sum_{k=0}^{K-1} \tau_k(\bm{s},t),
\end{equation*}
which, conditional on knot positions at resolution $K$ matching observation locations (cf. Section \ref{section:knotPlacement} for a detailed discussion of knot placement), we can re-write
\begin{equation}
    \widetilde{W}(\bm{s},t) = \sum_k \bm{b}_{k\ell_k}^\intercal(\bm{s},t) \bm{\eta}_{k\ell_k}. \label{eq:decomposition}
\end{equation}
The index $\ell_k$ depends on $\bm{s}$ and $t$, with $(\bm{s},t) \in R_{k\ell_k}$. The basis functions and coefficients are
\begin{align*} 
     \bm{b}_{k\ell}(\bm{s},t) & = \Cov[\delta_k(\bm{s},t), \bm{\Delta}_{k\ell} ], \hspace{0.5cm} (\bm{s},t) \in R_{k\ell} \\
     \bm{\Delta}_{k\ell} & = \{\delta_k(\bm{u},v) : (\bm{u},v)\in \bm{\mathcal{Q}}_k \cap R_{k\ell}\}, \\  
    \bm{\eta}_{k\ell} & \sim \text{N}(0, \bm{\Gamma}_{k\ell}), \\
    \bm{\Gamma}_{k\ell} & = \Var(\bm{\Delta}_{k\ell})^{-1}. 
\end{align*}

Although the notation is different, Eq. \ref{eq:decomposition} is equivalent to the original formulation in \citet{Katzfuss2017}, and all vectors $\bm{\eta}$ are mutually independent. In IS-MRA, we express the multivariate extension of Eq. \ref{eq:decomposition}
\begin{equation}
 \bm{\widetilde{W}} = \bm{F}(\bar{\bm{\Psi}}; \bm{\mathcal{Q}}) \bm{\eta}, \label{eq:wApprox}
\end{equation}
with $\bm{\mathcal{Q}}$ being the set of all knot positions, and $\bm{\eta}$ being the set of all $\bm{\eta}_{k\ell}$ vectors.

It follows from the definition of $\bm{\Gamma}_{k\ell}$ that $\bm{\eta}$ has a block diagonal covariance structure, which we denote $\bm{\Gamma}$, and we find that
\begin{equation}
 \Var(\bm{\widetilde{W}}) = \bm{F}(\bar{\bm{\Psi}}; \bm{\mathcal{Q}}) \bm{\Gamma} \bm{F}(\bar{\bm{\Psi}}; \bm{\mathcal{Q}})^\intercal. \label{eq:basicMRAapprox} 
\end{equation}
We stress that row $i$ of $\bm{F}(\bar{\bm{\Psi}}; \bm{\mathcal{Q}})$ is
\begin{equation*}
\bm{F}_{[i,\ ]}(\bar{\bm{\Psi}}; \bm{\mathcal{Q}}) = [\bm{b}_{01}^\intercal(\bm{s}_i, t_i), \bm{b}_{11}^\intercal(\bm{s}_i, t_i) \dots, \bm{b}_{1L_1}^\intercal(\bm{s}_i, t_i), \dots, \bm{b}_{K1}^\intercal(\bm{s}_i, t_i),\dots, \bm{b}_{KL_K}^\intercal(\bm{s}_i, t_i)],
\end{equation*}
and it follows that $\bm{F}(\bar{\bm{\Psi}}; \bm{\mathcal{Q}})$ is sparse, as $\bm{b}_{k\ell}^\intercal(\bm{s}_i, t_i)$ is non-zero only when $\ell = \ell_k$ such that $(\bm{s}_i, t_i) \in R_{k\ell_k}$. Moreover, $\bm{F}(\bar{\bm{\Psi}}; \bm{\mathcal{Q}})$ has number of rows and columns equal to the number of observations and number of knots across all regions, respectively. We can now see why $\bm{\widetilde{W}}$ is computationally tractable: its precision matrix is sparse because of the compact support of each basis function.

\subsection{Importance sampling for approximating hyperparameter posteriors}

Based on MRA, we obtain approximation
\begin{equation}
 \widetilde{\bm{Y}} \sim N[\bm{X} \bm{\beta} + \bm{F}(\bar{\bm{\Psi}}; \bm{\mathcal{Q}})\bm{\eta}, \zeta^2]. \label{eq:INLAy}
\end{equation}
Let $\bm{v}^{\intercal} = \{\bm{\beta}^{\intercal}, \bm{\eta}^{\intercal}\}$ denote the \textit{mean parameters}. We then rewrite Eq. \ref{eq:INLAy},
\begin{equation*}
 \widetilde{\bm{Y}} \mid \bm{v} \sim N[\bm{H}(\bar{\bm{\Psi}}; \bm{\mathcal{Q}}) \bm{v}, \zeta^2],
\end{equation*}
where
\[
 \bm{H}(\bar{\bm{\Psi}}; \bm{\mathcal{Q}}) = 
 \begin{bmatrix}
  \bm{X} & \bm{F}(\bar{\bm{\Psi}}; \bm{\mathcal{Q}})
 \end{bmatrix}.\]
The number of covariates is usually small compared to the total number of knots. As a result, although $\bm{X}$ is dense, $\bm{H}(\bar{\bm{\Psi}}; \bm{\mathcal{Q}})$ remains sparse.

The next steps are based on parts of the Integrated Nested Laplace Approximation (INLA) algorithm \citep{Rue2009, Martins2013}. Using Bayes law, we can show that
\begin{equation}
    p[\bm{\Psi} \mid \bm{y}] \propto \dfrac{p(\bm{\Psi})p(\bm{v} \mid \bm{\Psi})p[\bm{y} \mid \bm{v}, \bm{\Psi}]}{p[\bm{\bm{v}} \mid \bm{\Psi},\bm{y}]}. \label{eq:jointMarginal}
\end{equation}
Distribution $p[\bm{\Psi} \mid \bm{y}]$ is used to obtain marginals $p(\Psi_i \mid \bm{y})$ and $p[v_i \mid \bm{y}]$.

We assume independence between the prior distributions of $\bm{\beta}$ and $\bm{\eta}$, and as a result, $p(\bm{v} \mid \bm{\Psi})$ is equal to their product. As noted in Section \ref{section:MRA}, $\bm{\eta}$ is  multivariate Normal with mean $\bm{0}$ and covariance matrix $\bm{\Gamma}$. The likelihood $p[\bm{y} \mid \bm{v}, \bm{\Psi}]$ is multivariate Normal, with mean and covariance $\bm{H}(\bar{\bm{\Psi}}; \bm{\mathcal{Q}}) \bm{v}$ and $\zeta^2 \bm{I}$, respectively.

The distribution $p[\bm{\bm{v}} \mid \bm{\Psi},\bm{y}]$ is known as the \textit{full conditional}. It can be shown that when the likelihood is Normal, the full conditional is also Normal, with precision matrix \citep{Eidsvik2012}
\begin{equation}
    \Var(\bm{v} \mid \bm{\Psi}, \bm{y})^{-1} = \bm{Q} = 
    \begin{bmatrix}
        \bm{\Sigma}_{\beta} & \\
        & \bm{\Gamma}
    \end{bmatrix}^{-1} + \dfrac{1}{\zeta^2} \bm{H}(\bar{\bm{\Psi}}; \bm{\mathcal{Q}})^{\intercal} \bm{H}(\bar{\bm{\Psi}}; \bm{\mathcal{Q}}). \label{eq:Qmatrix}
\end{equation}

The full conditional mean is equal to $\bm{Q}^{-1} \bm{\xi}$, with $\bm{\xi} = (1/\zeta^2) \bm{H}(\bar{\bm{\Psi}}; \bm{\mathcal{Q}})^{\intercal} \bm{y}$ \citep{Eidsvik2012}. The precision matrix is therefore sparse. This results in a reasonable computational burden that can be controlled by adjusting the total number of knots and $K$, the depth of the decomposition of the spatiotemporal domain.

We cannot use Eq. \ref{eq:jointMarginal} directly to derive posteriors, as it is not standardised. We can however circumvent the issue with an importance sampling (IS) scheme. It involves obtaining $N_{IS}$ draws from a Gaussian proposal distribution denoted $p'(.)$, with mean equal to the mode of $p[\bm{\Psi} \mid \bm{y}]$, and covariance matrix equal to the inverse of the Hessian matrix computed numerically at the mode \citep{Rue2009}. Optimisation relies on the L-BFGS algorithm \citep{Liu1989}, with the gradient being estimated numerically by using a finite difference approximation.

Let,
\begin{equation*}
    p[\bm{\Psi} \mid \bm{y}] = \tilde{p}[\bm{\Psi} \mid \bm{y}]/c_I,
\end{equation*}
where $c_I$ is the unknown standardisation constant and $\widetilde{p}[\bm{\Psi} \mid \bm{y}]$ is the non-standardised distribution given in Eq. \ref{eq:jointMarginal}. We note that 
\begin{align*}
 \int_{\bm{\Psi}} p[\bm{\Psi} \mid \bm{y}] d\bm{\Psi} &= 1 \\
 &= \int_{\bm{\Psi}} \dfrac{p[\bm{\Psi} \mid \bm{y}]}{p'(\bm{\Psi})} p'(\bm{\Psi}) d\bm{\Psi} \\
 &\approx \dfrac{1}{N_{IS}}\sum_{i = 1}^{N_{IS}} \dfrac{p[\bm{\Psi}_i^\prime \mid \bm{y}]}{p'(\bm{\Psi}_i^\prime)} \\
 &= \dfrac{1}{N_{IS}} \sum_{i = 1}^{N_{IS}} \dfrac{\tilde{p}[\bm{\Psi}_i^\prime \mid \bm{y}]}{c_I p'(\bm{\Psi}_i^\prime)} \\ 
\end{align*}
where $\bm{\Psi}_i^\prime$ are values sampled from proposal distribution $p'(\bm{\Psi})$. It follows that,
\begin{equation*}
 c_I \approx \sum_{i = 1}^{N_{IS}} \dfrac{\tilde{p}[\bm{\Psi}_i^\prime \mid \bm{y}]}{N_{IS} p'(\bm{\Psi}_i^\prime)}.
\end{equation*}
The IS weight for the $i$'th draw from the proposal is therefore,
\begin{equation*}
 \omega_i = \dfrac{\tilde{p}[\bm{\Psi}_i^\prime \mid \bm{y}]}{c_{I} p'(\bm{\Psi_i}^\prime)}.
\end{equation*}
We obtain an empirical estimate of $p[\Psi_j \mid \bm{y}]$, $j = 1, \dots, n_{\Psi}$, by integrating $p[\bm{\Psi} \mid \bm{y}]$ based on the values sampled from the proposal distribution, that is,
\begin{equation*}
 p[\Psi_j \mid \bm{y}] \approx \delta(\Psi_j = \Psi_{i,j}^\prime) c_I \tilde{p}[\bm{\Psi} = \bm{\Psi}_i^\prime \mid \bm{y}], \hspace{10pt} i = 1, \dots, N_{IS},
\end{equation*}
where $\delta(.)$ is the delta function. We also have that,
\begin{equation*}
 p[v_j \mid \bm{y}] \approx \sum_{i = 1}^{N_{IS}} p[v_j \mid \bm{\Psi}_i^\prime,\bm{y}] \omega_i.
\end{equation*}
Note that since the full conditional is multivariate normal, $p[v_j \mid \bm{\Psi},\bm{y}]$ can be readily obtained. Values obtained through the IS algorithm can let us readily derive empirical estimates of the moments and quantiles of the marginal distributions. However, as extreme percentiles of those distributions may be hard to estimate accurately, IS-MRA uses a skew-normal approximation to obtain credible interval bounds, with parameters estimated with the method of moments. If the sampling distribution has too much skewness however, method of moments estimation fails, and the empirical quantiles are used instead. 

\subsection{Spatiotemporal prediction}

Let $y^P_k$ denote the predicted responses at coordinate $(\bm{s}_k^P, t_k^P)$. We aim to obtain moments of the posterior predictive distribution $p[\bm{y}^P \mid \bm{y}]$. The posterior predictive distribution itself is computationally intractable, as it involves the dense covariance matrix of $\bm{W}$. We define $\bm{F}_P(\bar{\bm{\Psi}}; \bm{\mathcal{Q}})$ similarly to $\bm{F}(\bar{\bm{\Psi}}; \bm{\mathcal{Q}})$, but with the spatiotemporal coordinates $(\bm{s}_k^P, t_k^P)$ in place of $(\bm{s}_i, t_i)$, and 
\[
\bm{H}_P(\bar{\bm{\Psi}}; \bm{\mathcal{Q}}) =
\begin{bmatrix}
    \bm{X}^P & \bm{F}_P(\bar{\bm{\Psi}}; \bm{\mathcal{Q}})
\end{bmatrix},
\]
where $\bm{X}^P$ has row $k$ with covariates at $(\bm{s}_k^P, t_k^P)$. We have 
\begin{equation*}
    E[\bm{Y}^P \mid \bm{y}, \bm{\Psi}, \bm{v}] =  \bm{H}_P(\bar{\bm{\Psi}}; \bm{\mathcal{Q}}) \bm{v}
\end{equation*}
and
\begin{align*}
    E[\bm{Y}^P \mid \bm{y}, \bm{\Psi}] &= E_{\bm{v}}\{E[\bm{Y}^P \mid \bm{y}, \bm{\Psi}, \bm{v}]\} \\
    &= E_{\bm{v}}[\bm{H}_P(\bar{\bm{\Psi}}; \bm{\mathcal{Q}}) \bm{v}] \\
    &= \bm{H}_P(\bar{\bm{\Psi}}; \bm{\mathcal{Q}}) \bm{Q}^{-1}\bm{\xi}.
\end{align*}
We then obtain
\begin{align*}
 E[\bm{Y}^P \mid \bm{y}] &=  E_{\bm{\Psi}}\{E[\bm{Y}^P \mid \bm{y}, \bm{\Psi}]\} \\
 &\approx \dfrac{1}{N_{IS}}\sum_{i = 1}^{N_{IS}} \bm{H}_P(\bm{\Psi}_i^\prime; \bm{\mathcal{Q}}) \bm{Q}^{-1}_i \bm{\xi}_i \omega_i, 
\end{align*}
where $\bm{Q}_i$ and $\bm{\xi}_i$ are the $\bm{Q}$ matrix and $\bm{\xi}$ vector obtained conditional on $\bm{\Psi} = \bm{\Psi}_i^\prime$. To obtain marginal variances $\Var[Y_j^P \mid \bm{y}]$, $j = 1, \dots, n_P$, we first note that
\begin{equation*}
 \Var[\bm{Y}^P \mid  \bm{y}, \bm{\Psi}] = \Var_{\bm{v}}\{E[\bm{Y}^P \mid \bm{y}, \bm{\Psi}, \bm{v}]\} + E_{\bm{v}}\{\Var[\bm{Y}^P \mid \bm{y}, \bm{\Psi}, \bm{v}]\}.
\end{equation*}
$\Var[\bm{Y}^P \mid \bm{y}, \bm{\Psi}, \bm{v}]$ is equal to $\zeta^2 \bm{I}$ , which corresponds to the value of the second term. We then note that,
\begin{align*}
    \Var\{E[\bm{Y}^P \mid \bm{y}, \bm{\Psi}, \bm{v}]\} &= \Var\{\bm{H}_P(\bar{\bm{\Psi}}; \bm{\mathcal{Q}})\bm{v}\} \\
    &= \bm{H}_P(\bar{\bm{\Psi}}; \bm{\mathcal{Q}}) \bm{Q}^{-1} \bm{H}_P(\bar{\bm{\Psi}}; \bm{\mathcal{Q}})^{\intercal}
\end{align*}
Again, we have that,
\begin{equation*}
    \Var[\bm{Y}^P \mid  \bm{y}] = \Var_{\bm{\Psi}}\{E[\bm{Y}^P \mid \bm{y}, \bm{\Psi}]\} + E_{\bm{\Psi}}\{\Var[\bm{Y}^P \mid \bm{y}, \bm{\Psi}]\}.
\end{equation*}
We can once again use our IS weights to estimate the first and second terms, as both are expectations with respect to $\bm{\Psi}$. We provide additional details in Appendix A.


\subsection{Knot placement} \label{section:knotPlacement}

At resolution $K$, the finest resolution, IS-MRA places knots at observation locations. The total number of knots at resolution $r, r = 0, \dots, K-1$ is $M_0 J^r$, where, by default, $M_0$ and $J$ take values $20$ and $2$ respectively. Knot placement at those resolutions is guided by two principles. First, we want to place some knots at prediction locations, as posterior predictive distributions are based on an interpolation scheme involving values computed at the knots. Then, based on recommendations in \citet{Katzfuss2017}, we prefer knot positions to be close to the subregion boundaries.

For simplicity, let us assume that each subregion contains $M_0$ knots. We process subregions from coarsest to finest, and only one knot can be placed at any given location. It follows that once a knot has been placed at a given prediction location, that location becomes \textit{unavailable} and cannot be selected again. If a subregion contains fewer than $M_0$ available prediction locations, a first batch of knots is placed at those locations. Then, the algorithm places the missing knots on the edges of a rectangular prism nested within that subregion. It also adds a small amount of jittering to prevent knots from perfectly overlapping. Otherwise, if the region contains more than $M_0$ available prediction locations, the knots are placed by sampling uniformly at random without replacement $M_0$ values from the set of available prediction locations. Alternatively, IS-MRA can simply place knots uniformly at random.

\subsection{Software}

An R package to fit IS-MRA is available at \url{https://github.com/villandre/MRAinla/}. The software is written in R and C++, with the interface between the two relying on the \textit{Rcpp} \citep{Eddelbuettel2011, Eddelbuettel2013, Eddelbuettel2017} and \textit{RcppEigen} \citep{Bates2013} packages. \textit{RcppEigen} relies on the \textit{Eigen} package \citep{Eigen}. 


\begin{figure}[ht]
    \centering
    \includegraphics[width = 6cm]{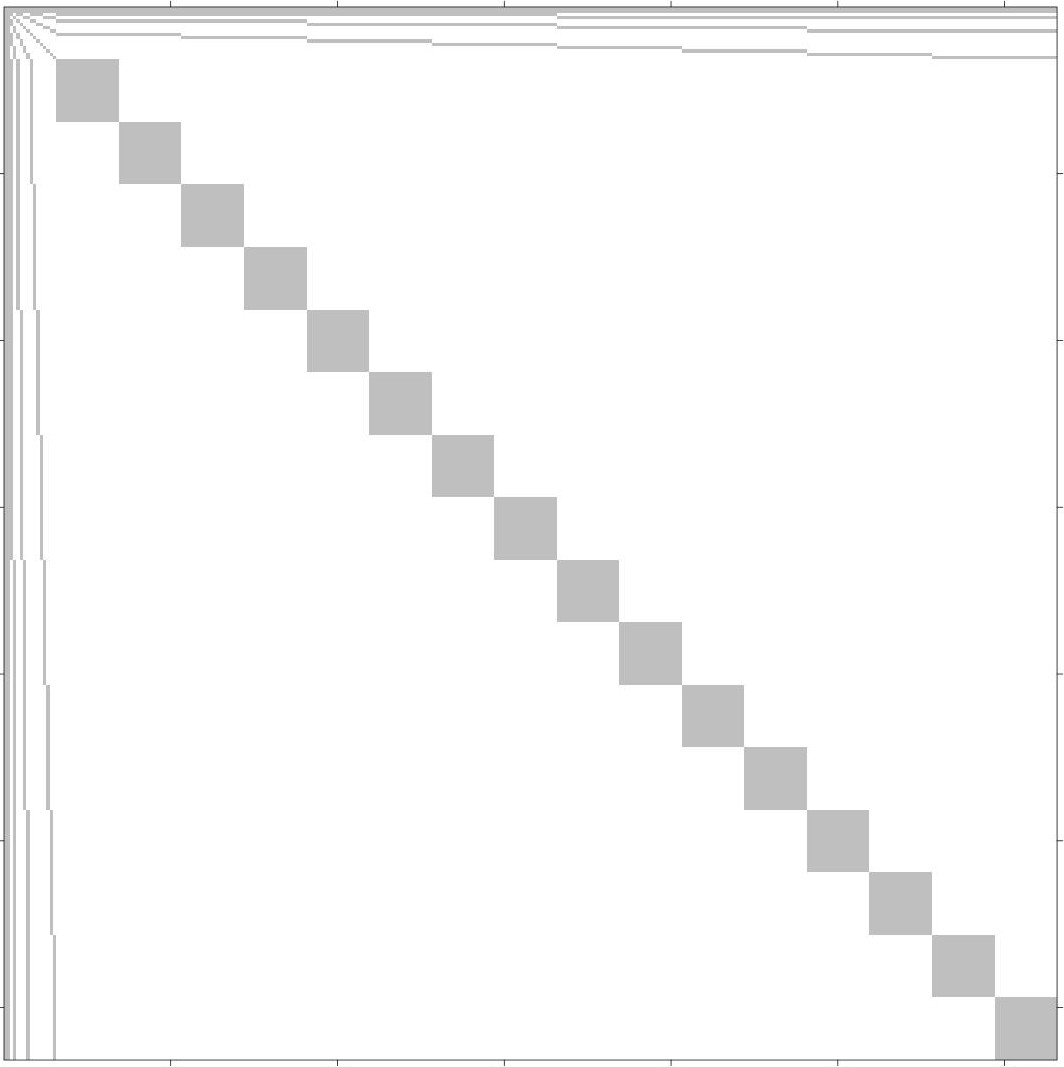}
    \caption{\textbf{Example of precision matrix $\bm{Q}$.} Grey and white tiles represent non-zero-valued and zero-valued elements, respectively.}
    \label{fig:matrixSparsity}
\end{figure}

The main computational challenge results from expressions like $\bm{Q}^{-1}\bm{x}$, which appear, for example, in the full conditional mean, and the posterior predictive means and variances. Since the $\bm{Q}$ matrix is sparse, its Cholesky ($LDL^\intercal$) decomposition can be obtained.
We present in Figure \ref{fig:matrixSparsity} a graphical representation of a typical $\bm{Q}$ matrix. It comprises a large number of blocks, with the lower-right section, a block-diagonal submatrix, being by far the largest. It corresponds to the knots used at the finest resolution, which are placed at observation locations. We stress that the sparsity pattern is independent of the simulated responses, and assumed parameter and hyperparameter values: it depends solely on spatiotemporal coordinates. Further, all applications of IS-MRA will produce a similar pattern. We take advantage of the independence between the structure of the matrix and the hyperparameter values to obtain considerable computational gains in the IS step. Indeed, the sparsity structure analysis used in the computation of the Cholesky decomposition need be performed only once, and the result is re-used for all IS points.

The creation of large sparse matrices is also time-consuming. After initialisation, $\bm{H}(\bar{\bm{\Psi}}; \bm{\mathcal{Q}})$ is updated by iterating through its non-zero elements only. The computation of the posterior predictive variances is also fairly demanding, as it involves solving as many linear systems as observations in the sample. Thanks to \texttt{openMP} parallelisation, we are able to complete that step reasonably quick. For this reason, we strongly recommend running the software on a machine with \texttt{openMP} support enabled. We provide additional information on the software in Appendix A.

\section{Results}

We investigate the properties of IS-MRA through a real data analysis coupled with a validation exercise, and a simulation study. We ran all computations on a machine equipped with an Intel(R) Xeon(R) W-2145 CPU @ 3.70GHz CPU (16 cores) and 188GB of memory. Datasets and output files, and scripts used to produce the results can be found, respectively, at \url{https://github.com/villandre/dataFilesForAnalyses} and \url{https://github.com/villandre/ISMRAanalyses}. We parallelised the computation of posterior predictive distributions on eight cores.

\subsection{Application} \label{section:application}

The main analysis focuses on a dataset consisting of $1,116,319$ daytime LST observations, derived from MODIS spectroradiometry data collected between May 25 and May 31, 2012 in the western part of the state of Maharashtra, India. On each day,
we have two LST datasets, one from Terra and one from Aqua, which fly over the region around 11AM and 1PM, respectively. In our analyses, for simplicity, we consider only one set of measurements for each day, that is, of the two, we keep the one with the fewest missing values. The main objective is to impute the $81,574$ LSTs missing over non-oceanic tiles on May 28. We consider four covariates: land cover, elevation, satellite, i.e. Terra or Aqua, and day of observation. We obtained land cover values from the Terra and Aqua combined MODIS Land Cover Type (MCD12Q1) Version 6 data product \citep{MODIS_LandCover}, and we used elevation estimates from the ASTER Global Digital Elevation Model Version 3 \citep{MODIS_ASTER}. We include the satellite covariate to reflect a potential discrepancy in temperatures recorded by Terra and Aqua, due to the different fly over times. Finally, we create a day-of-observation covariate using dummy variables. It is analogous to a ``cluster-specific intercept'', that is, it provides a region-wide adjustment to the mean temperature observed on any given day.

\begin{table}[p] 
    \begin{center}
        \resizebox{0.98\textwidth}{!}{
            \begin{tabular}{lrrrr}                
                \toprule\multicolumn{1}{l}{}&\multicolumn{1}{c}{\textbf{Mean}}&\multicolumn{1}{c}{\textbf{StdDev}}&\multicolumn{1}{c}{\textbf{CI-2.5\%}}&\multicolumn{1}{c}{\textbf{CI-97.5\%}}\tabularnewline
                \hline
                {\bfseries Elevation, satellite}&&&&\tabularnewline
                ~~Elevation (per 1000m)& $-1.091$ & $0.019$ & $-1.128$ & $-1.055$ \tabularnewline
                ~~Satellite: Aqua&$ 0.793$&$4.472$&$ -7.905$&$ 9.490$\tabularnewline
                \midrule
                {\bfseries Land cover}&&&&\tabularnewline
                ~~Evergreen Needleleaf Forests&$ 1.090$&$0.108$&$  0.880$&$ 1.301$\tabularnewline
                ~~Evergreen Broadleaf Forests&$ 0.979$&$0.025$&$  0.931$&$ 1.026$\tabularnewline
                ~~Deciduous Broadleaf Forests&$ 1.181$&$0.022$&$  1.139$&$ 1.223$\tabularnewline
                ~~Mixed Forests&$ 1.087$&$0.022$&$  1.044$&$ 1.131$\tabularnewline
                ~~Closed Shrublands&$ 2.097$&$0.129$&$  1.846$&$ 2.347$\tabularnewline
                ~~Open Shrublands&$ 1.448$&$0.060$&$  1.332$&$ 1.565$\tabularnewline
                ~~Woody Savannas&$ 1.329$&$0.028$&$  1.276$&$ 1.383$\tabularnewline
                ~~Savannas&$ 1.393$&$0.018$&$  1.358$&$ 1.427$\tabularnewline
                ~~Grasslands&$ 1.506$&$0.018$&$  1.472$&$ 1.540$\tabularnewline
                ~~Permanent Wetlands&$ 0.705$&$0.020$&$  0.666$&$ 0.744$\tabularnewline
                ~~Croplands&$ 1.588$&$0.018$&$  1.554$&$ 1.623$\tabularnewline
                ~~Urban and Built-up Lands&$ 1.623$&$0.019$&$  1.587$&$ 1.659$\tabularnewline
                ~~Cropland/Natural Vegetation Mosaics&$ 1.430$&$0.021$&$  1.389$&$ 1.470$\tabularnewline
                ~~Non-Vegetated Lands&$ 0.667$&$0.028$&$  0.612$&$ 0.721$\tabularnewline
                \midrule
                {\bfseries Time}&&&&\tabularnewline
                ~~May 26&$-2.980$&$4.472$&$-11.677$&$ 5.717$\tabularnewline
                ~~May 27&$-0.033$&$4.472$&$ -8.731$&$ 8.664$\tabularnewline
                ~~May 28&$-6.319$&$0.071$&$ -6.458$&$-6.180$\tabularnewline
                ~~May 29&$ 1.155$&$4.472$&$ -7.542$&$ 9.853$\tabularnewline
                ~~May 30&$-3.438$&$0.078$&$ -3.590$&$-3.286$\tabularnewline
                ~~May 31&$ 2.650$&$4.472$&$ -6.047$&$11.348$\tabularnewline
                \midrule
                {\bfseries Hyperparameters}&&&&\tabularnewline
                ~~Spatial range ($\rho$) &$ 5.660$&$0.014$&$  5.632$&$ 5.688$\tabularnewline
                ~~Temporal range ($\phi$)&$ 3.601$&$0.018$&$  3.567$&$ 3.635$\tabularnewline
                ~~Std. dev. ($\sigma$)&$ 4.140$&$0.013$&$  4.114$&$ 4.166$\tabularnewline               
                \bottomrule
            \end{tabular}
        }
    \end{center}    
    \caption{\textbf{Mean and standard deviation of posteriors with credible intervals}. Water (land cover = 0 in UMD classification) is the reference category for land cover. ``Time = May 25'' is the reference category for the time parameters, and ``Terra'' is the reference for the ``Satellite: Aqua'' parameter.} \label{tab:hyperAndMeanTable}    
\end{table}

Mean, standard deviations, and credible intervals for the fixed effects and log-hyperparameter posteriors are given in Table \ref{tab:hyperAndMeanTable}.  The spatial range hyperparameter translates to correlations of $0.962$ at one kilometer, and $0.19$ at ten kilometers. For time, we have instead a correlation of $0.758$ after one day, and $0.143$ after a week. We found that the effect of land cover was usually modest, but supported by narrow credible intervals. We observed the most compelling effects for closed shrublands (Mean = $2.097$, SD = $0.129$) and urban and built up environments (Mean = $1.623$, SD = $0.019$). We found the day of observation covariate however to be fairly influential. For example, the decrease in the posterior means due to time on May 28 was $6.319$ degrees Celsius (SD = $0.071$), with May 25 serving as the reference. When the model could not reliably identify a mean difference, it produced a fairly high standard deviation for the associated posterior, around $4.472$.

\begin{figure}[p] 
    \begin{subfigure}[b]{0.48\textwidth}
    \centering
    \includegraphics[width = 6.4cm]{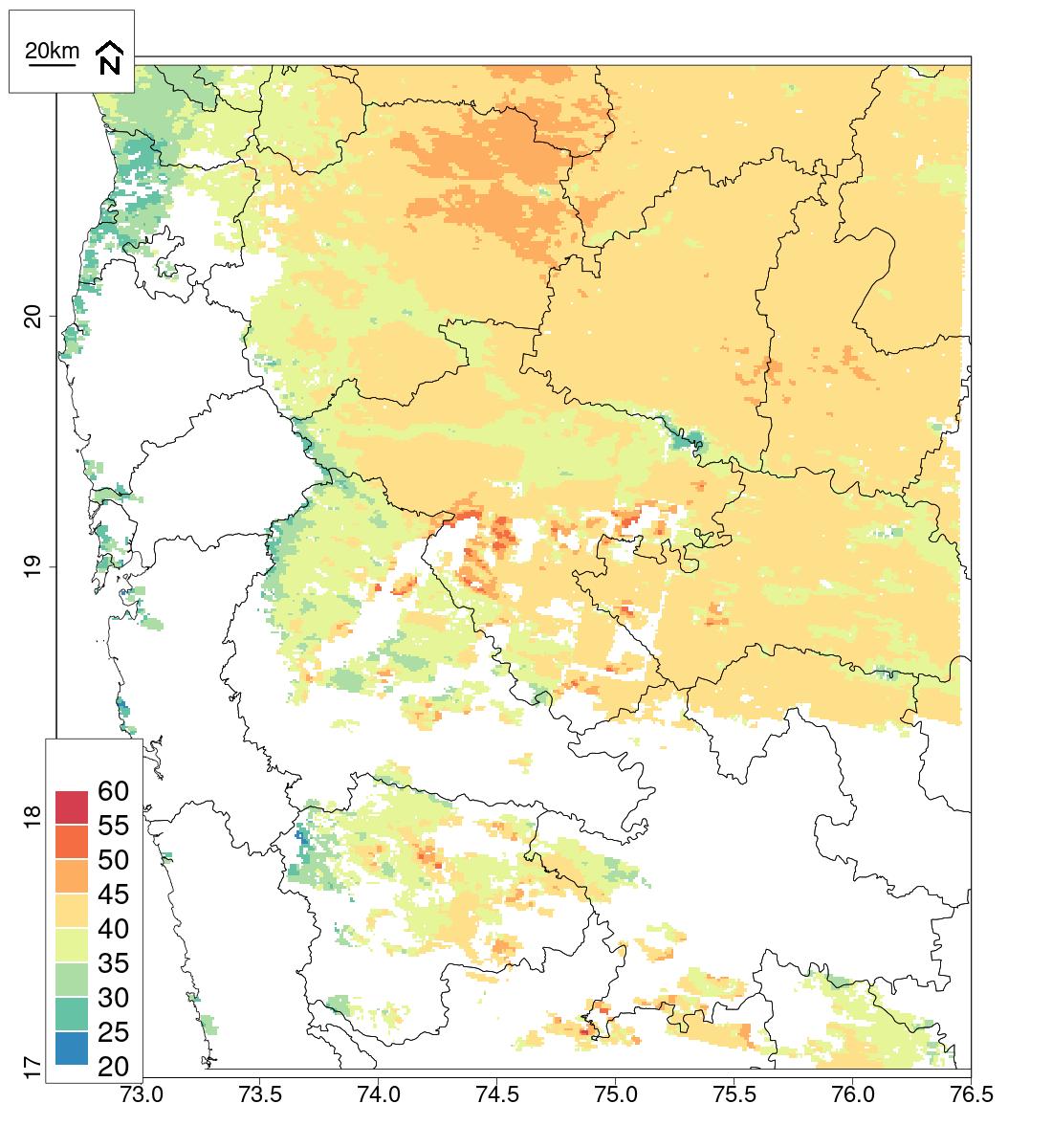}
    \caption{Observed data} \label{fig:mainPredictionsTrainingMay28}
    \end{subfigure}
    ~
    \begin{subfigure}[b]{0.48\textwidth}
    \centering
    \includegraphics[width = 6.4cm]{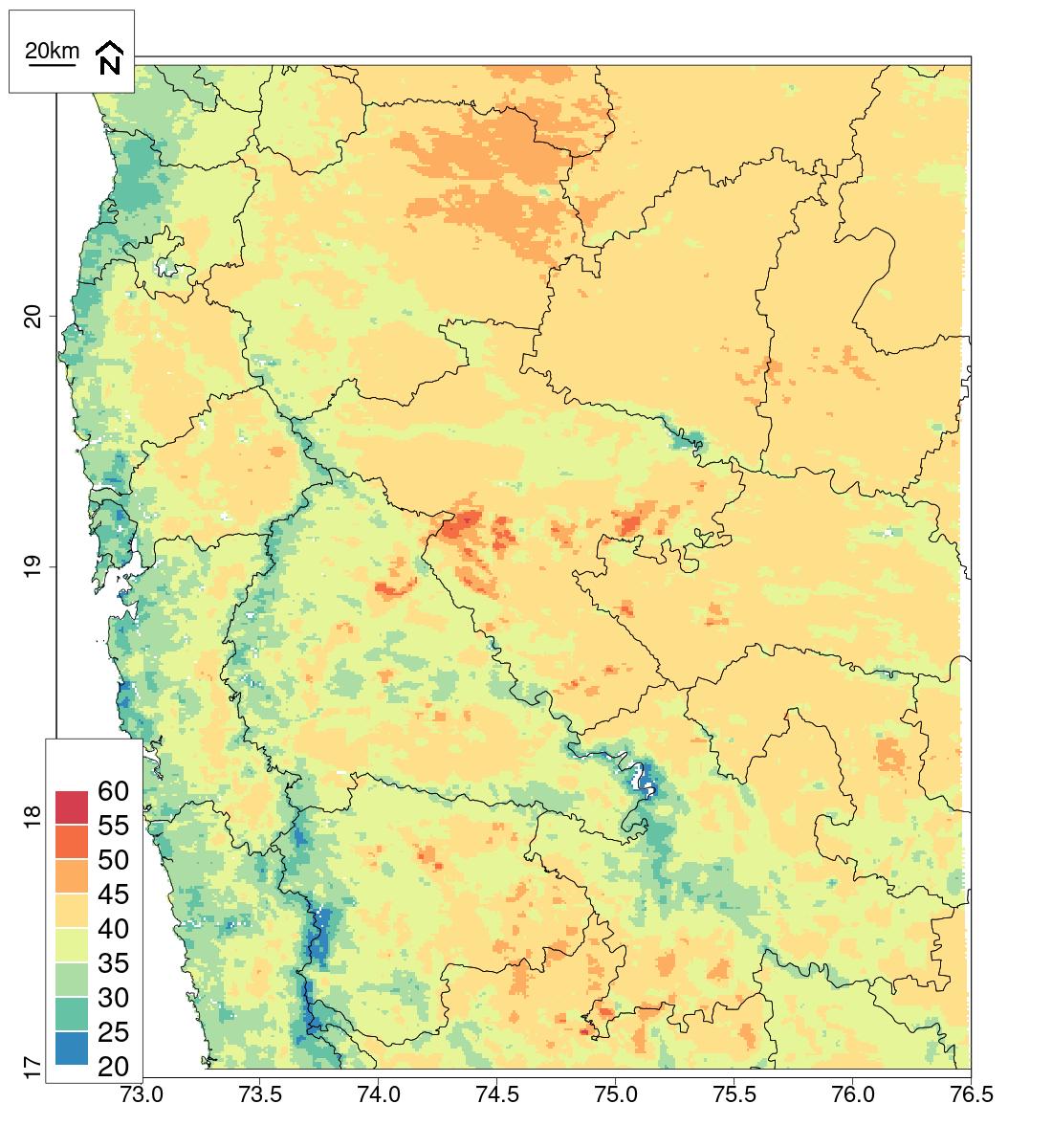}
    \caption{Observed data and predictions} \label{fig:mainPredictionsFitted}
    \end{subfigure}
    \begin{subfigure}[b]{0.48\textwidth}
    \centering
    \includegraphics[width = 6.4cm]{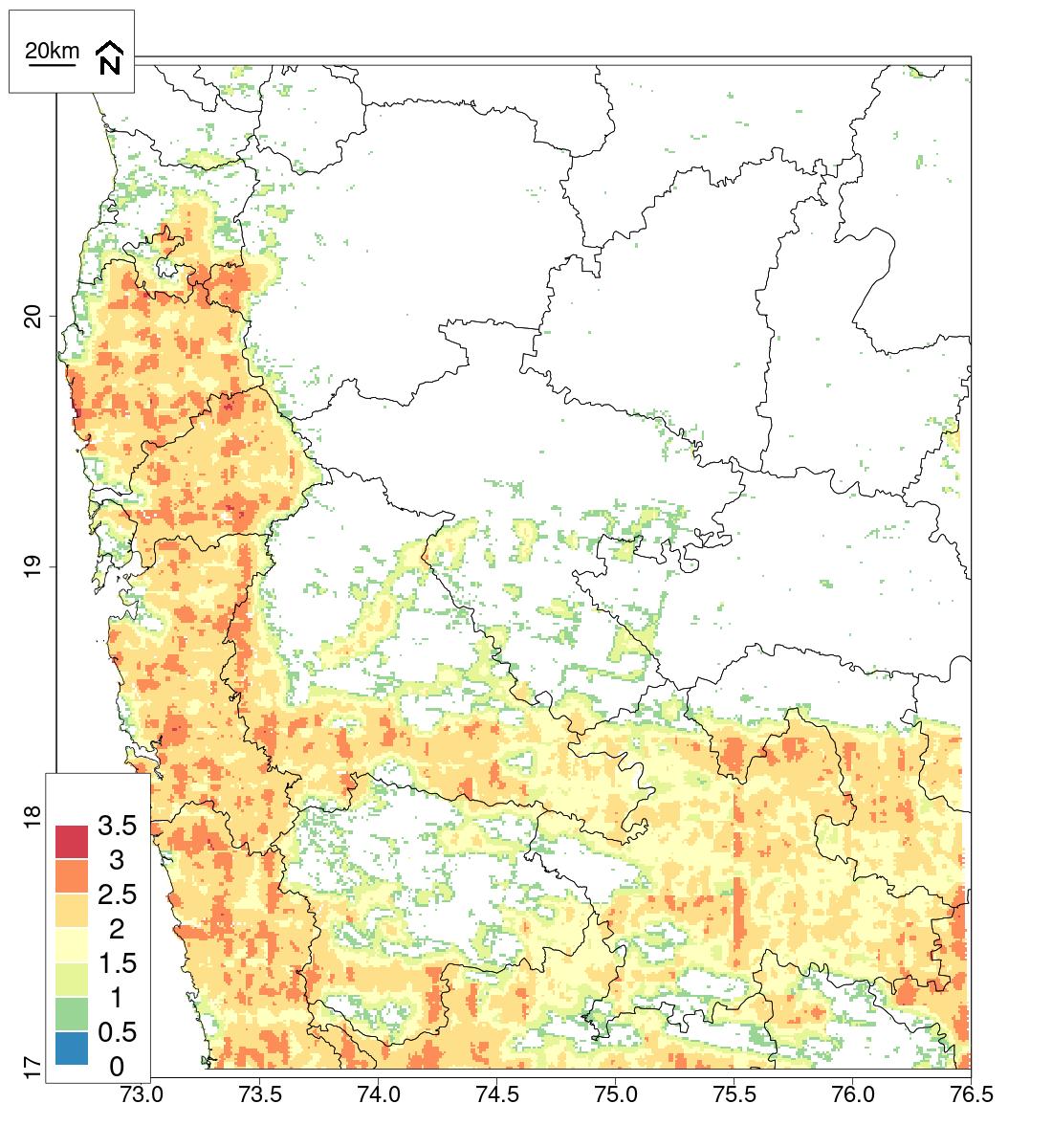}
    \caption{Posterior standard deviations} \label{fig:mainPredictionsSDs}
    \end{subfigure}    
    \caption{\textbf{Observed LST data, predictions $E[Y^P \mid \bm{y}]$, and standard deviations $\StdDev[Y^P \mid \bm{y}]$.} In b), $259$ predicted values falling outside the range of observed LSTs on May 28 were omitted before plotting to better highlight similarities with a) and variation among displayed values.}
    \label{fig:mainPredictionsMeanAndSDs}
\end{figure}

Fig. \ref{fig:mainPredictionsMeanAndSDs} shows the observed data for May 28, as well as the prediction surface from IS-MRA. The predictions form a realistic pattern overall, with lower predicted temperatures near the coast even though there are few observed data. The surface is not uniformly smooth: several small vertical breaks are present in the lower-middle section of the prediction map. Those breaks result from the partitioning of the domain \citep{Katzfuss2017b}. Further, we observe lower standard deviations in posterior predictive distributions for locations closer to where data were available on May 28. This is due to the higher spatial correlation inherent to LST data. The largest standard deviations take values slightly under $3.2$ degrees Celsius.

The method did output a small number of more extreme values. Out of the $81,574$ predicted values, $77$ are below the range of the sampled LSTs by at least one degree, while none are above it. The lowest prediction is $4.41$ degrees Celsius, and is for a coastal tile whose center is located at ($18.888$, $72.905$), just south of Mumbai. The second most extreme prediction, at $14.76$ degrees Celsius, is for another coastal tile, located immediately north of the first one ($18.896$, $72.908$).

\subsection{Validation}

We validate model predictions with a second dataset, whose observations were collected between May 18, 2012 and May 24, 2012 across a subset of the region considered in the main analysis. We split that dataset into training and test sets, comprising $108,079$ and $13,234$ observations, respectively. We create the test set by holding out observations falling under a simulated cloud cover on May 21. To obtain a realistic missingness pattern, we reproduce the cloud cover recorded on May 28. Assuming the same model as before, we then use IS-MRA to estimate the posterior predictive means, and we finally compute prediction errors. We provide more information regarding tuning parameters, priors, hyperpriors, and the recursive domain splitting scheme in Appendix B.

\begin{figure}[p]
    \begin{subfigure}[b]{0.48\textwidth}
    \centering
    \includegraphics[width = 5.8cm]{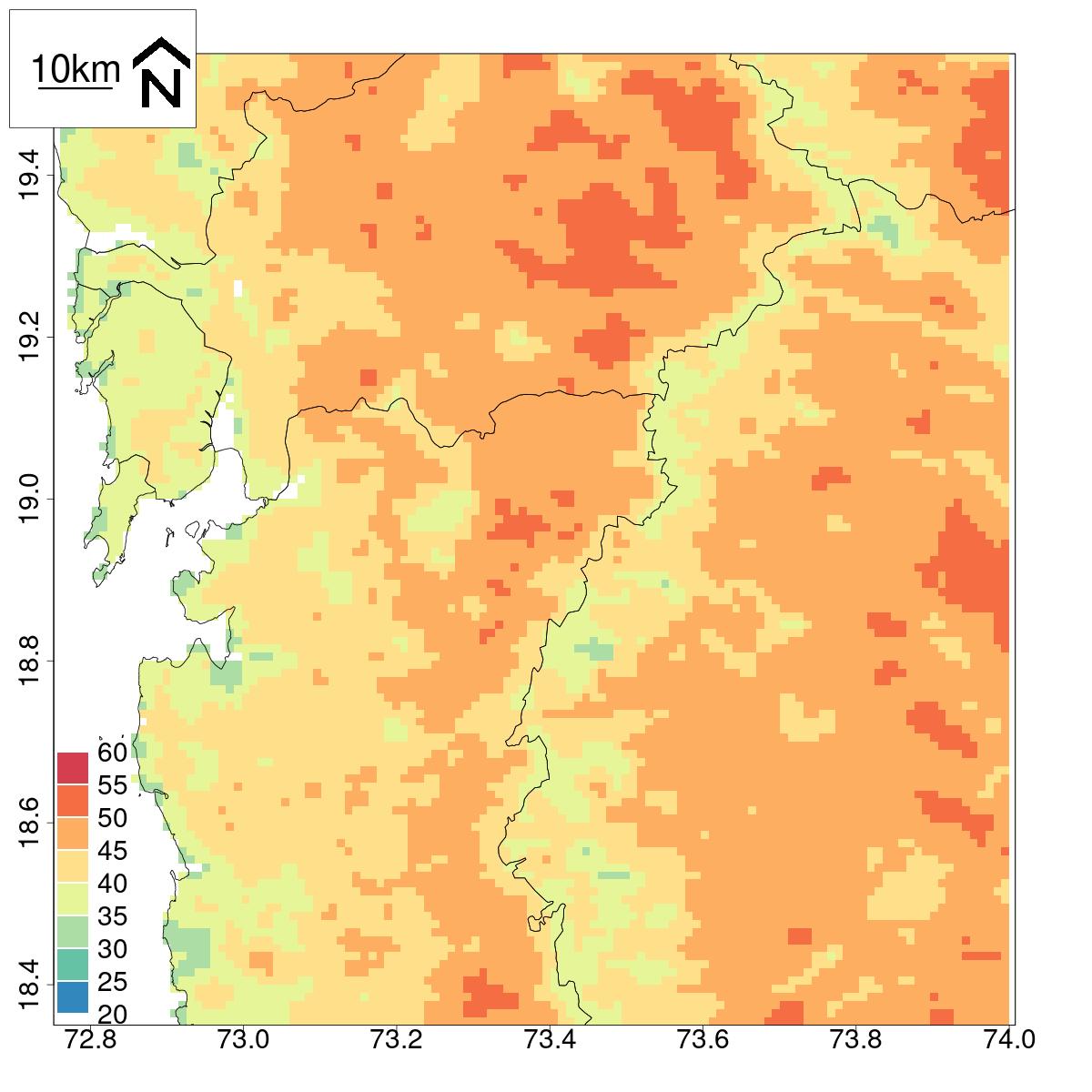}
    \caption{Observed LST data} \label{fig:validationOriData}
    \end{subfigure}
    ~
    \begin{subfigure}[b]{0.48\textwidth}
    \centering
    \includegraphics[width = 5.8cm]{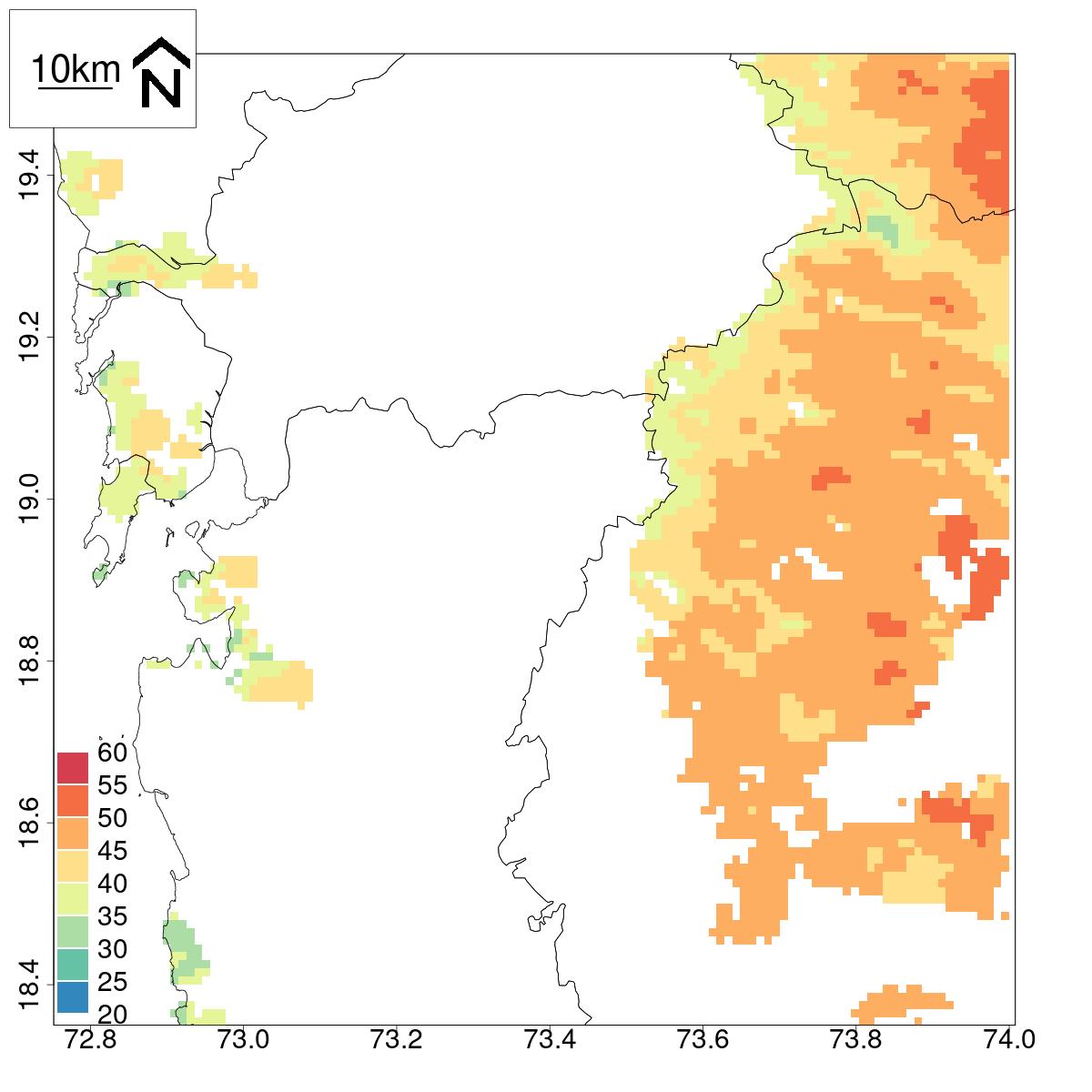}
    \caption{Training data} \label{fig:validationTrainingData}
    \end{subfigure}
    \begin{subfigure}[b]{0.48\textwidth}
    \centering
    \includegraphics[width = 5.8cm]{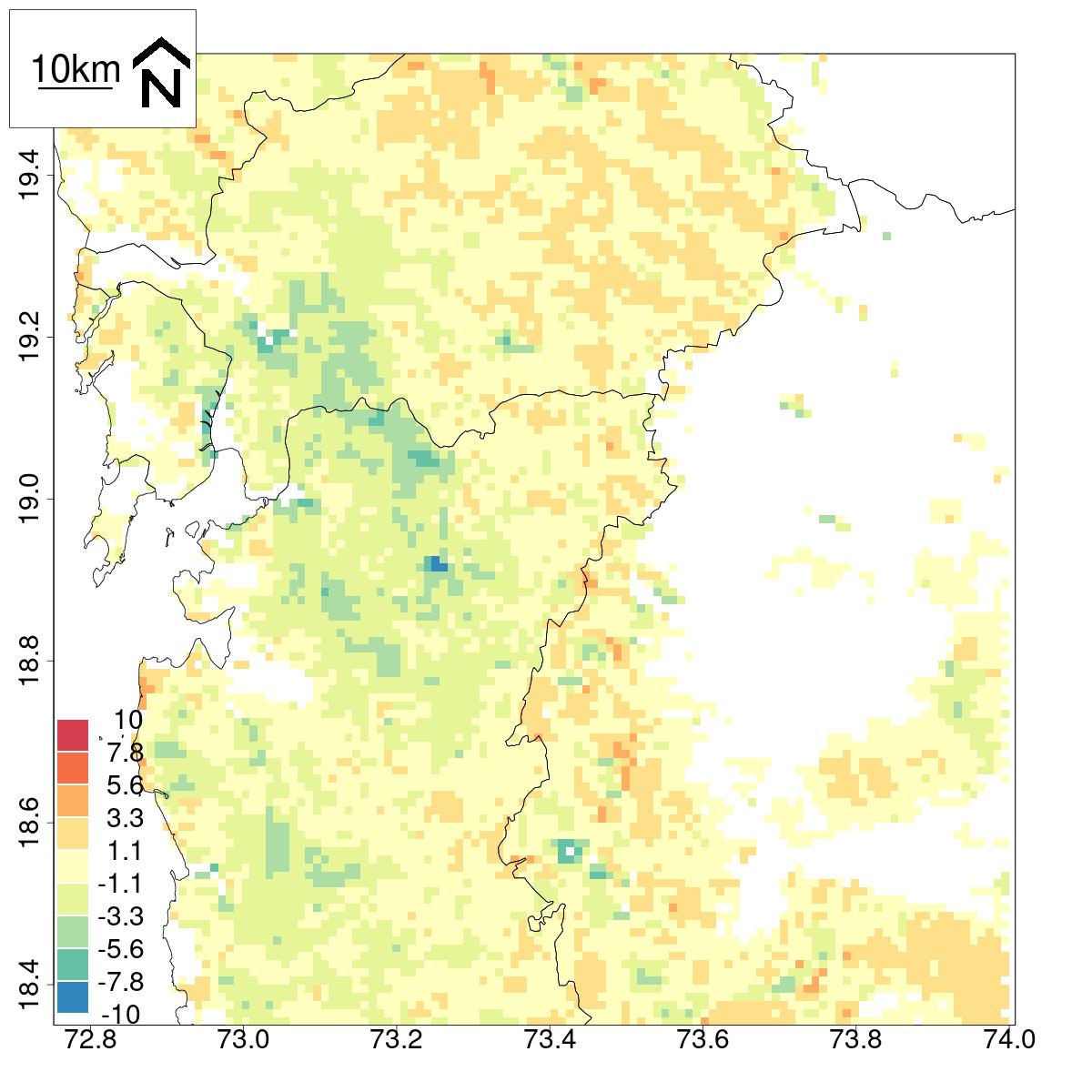}
    \caption{Prediction errors from IS-MRA} \label{fig:validationErrorsINLAMRA}
    \end{subfigure}
    \begin{subfigure}[b]{0.48\textwidth}
    \centering
    \includegraphics[width = 5.8cm]{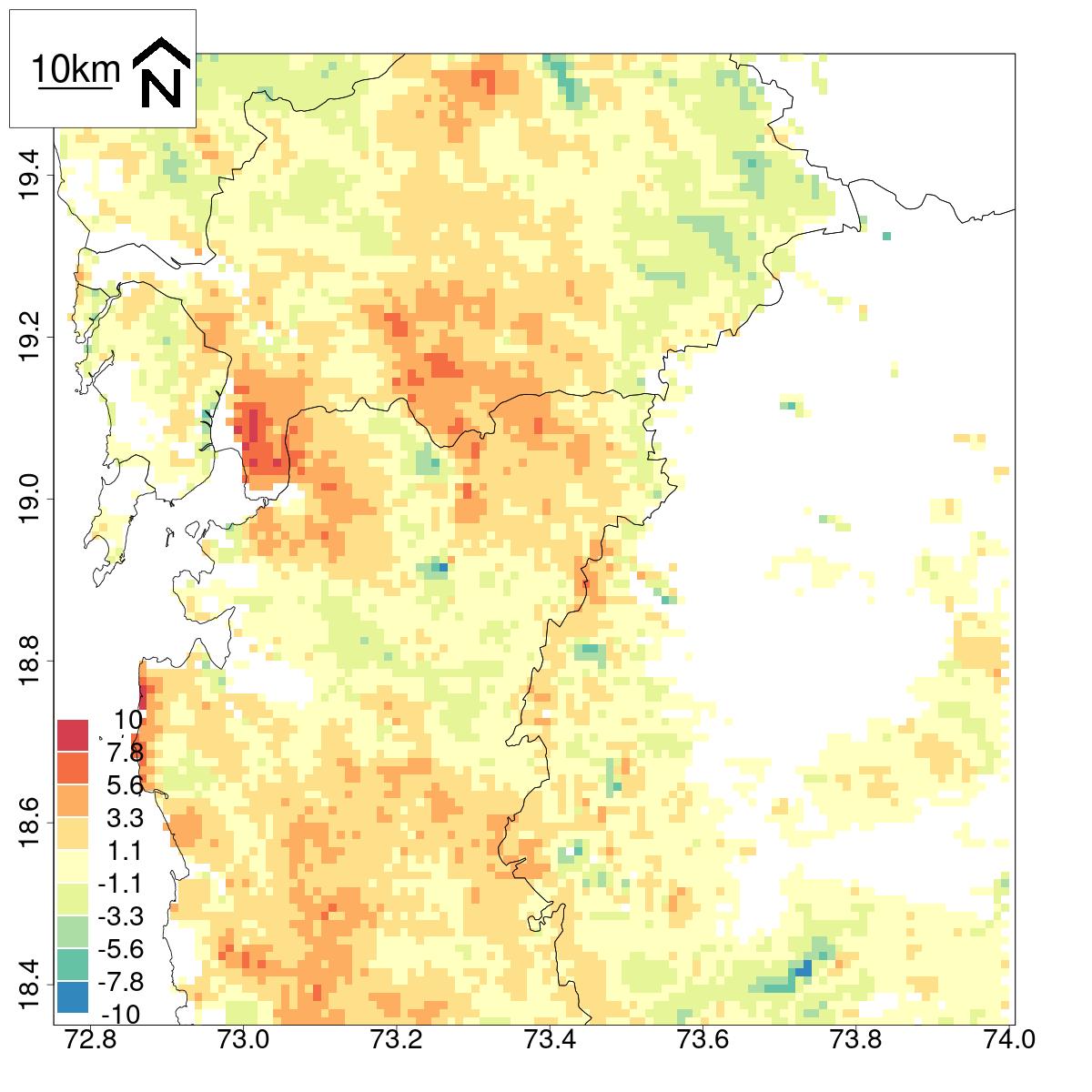}
    \caption{Prediction errors from INLA-SPDE} \label{fig:validationErrorsSPDE}
    \end{subfigure}
    \caption{\textbf{Validation results comparing IS-MRA and INLA-SPDE} The training data on May 21 is obtained by masking the observed data with the cloud pattern on May 28. Prediction errors for IS-MRA and INLA-SPDE are equal to observed values minus posterior means.}
    \label{fig:validationData}    
\end{figure}

Fig. \ref{fig:validationOriData} displays the LST measurements recorded on May 21, and Fig. \ref{fig:validationTrainingData}, the values we subtracted to reproduce the cloud cover observed on May 28. Fig. \ref{fig:validationErrorsINLAMRA} shows the errors obtained by subtracting the observed values from the corresponding posterior predictive means. As expected, there is visible spatial structure in the errors we obtained, which stretch from $-10.16$ to $5.98$ degrees Celsius. The largest absolute difference is for a water tile on the Morbe lake ($73.250, 18.913$). The mean squared prediction error (MSPE) and median squared prediction error (MedSPE) are $2.94$ and $1.13$, respectively, which indicates that the available data were reasonably informative. Further, $90$\% of absolute errors are under $2.83$. We suspect the more extreme values result from the selected model's difficulties in predicting accurately on or next to water bodies.

We compare IS-MRA to INLA-SPDE \citep{Lindgren2011}, the latter method also making use of a convenient basis representation for the GRF to propose a sparse precision matrix. We used functions in the INLA package (\url{www.rinla.org}) to fit INLA-SPDE. As expected, INLA-SPDE also did well, with MSPE and MedSPE equal to $4.94$ and $2.02$, respectively. We map prediction errors in Fig. \ref{fig:validationErrorsSPDE}, on the same scale as in Fig. \ref{fig:validationErrorsINLAMRA} to facilitate comparison. The figure highlights that INLA-SPDE had a tendency to overestimate observed LSTs, while IS-MRA had a tendency to underestimate them instead. Both approaches had comparable computational requirements in time and memory, with computational times depending strongly on the number of basis functions used.

\subsection{Simulation study}

We simulated $100$ datasets of size $6,546$, mimicking LST observations collected on three consecutive days in a region bounded by longitudes $73.2$ and $73.4$ and latitudes $18.6$ and $18.8$. We exclude from each dataset a block of observations bounded by longitudes $73.25$ and $73.35$ and latitudes $18.65$ and $18.75$, resulting in a test set of size $546$. We consider three covariates: land cover and elevation, which we obtain the same way as before \citep{MODIS_LandCover, MODIS_ASTER}, and day of observation, expressed with dummy variables. In order to simulate LST values, we used the means of the marginal posterior distributions for parameters and hyperparameters presented in Section \ref{section:application}. A plot of one of the simulated training datasets can be found in Appendix C. 

IS-MRA and INLA-SPDE assume different parameterisations for hyperparameters, and it is not possible to use priors which match perfectly. We use the default INLA parameterisation, under which a prior is assigned to 
\begin{equation}
    \theta_{INLA} = \log\left(\dfrac{1+\exp(-1/\phi)}{1 - \exp(-1/\phi)}\right), \label{eq:INLAtimeRange}
\end{equation}
and derive matching moments for $\theta_{INLA}$ by simulating random numbers from the prior assigned to $\phi$ in IS-MRA. We then assign $\theta_{INLA}$ a normal prior with the mean and standard deviation we obtain.  

In order to make a fair comparison, we build the mesh in INLA-SPDE with a number of nodes giving the precision matrix for the GRF approximately as many non-zero elements as the $\bm{Q}$ matrix in IS-MRA. We end up with both of them having approximately $1.4$ million non-zero entries. To better highlight the effects of imposing very high sparsity on $\bm{Q}$ in IS-MRA, we also consider a case where the number of non-zero elements is $2.8$ million, which we achieve by decreasing $K$ from $6$, the value used in the original scenario, to $4$.

\begin{figure}[ht] 
    \centering
    \includegraphics[width = 8cm]{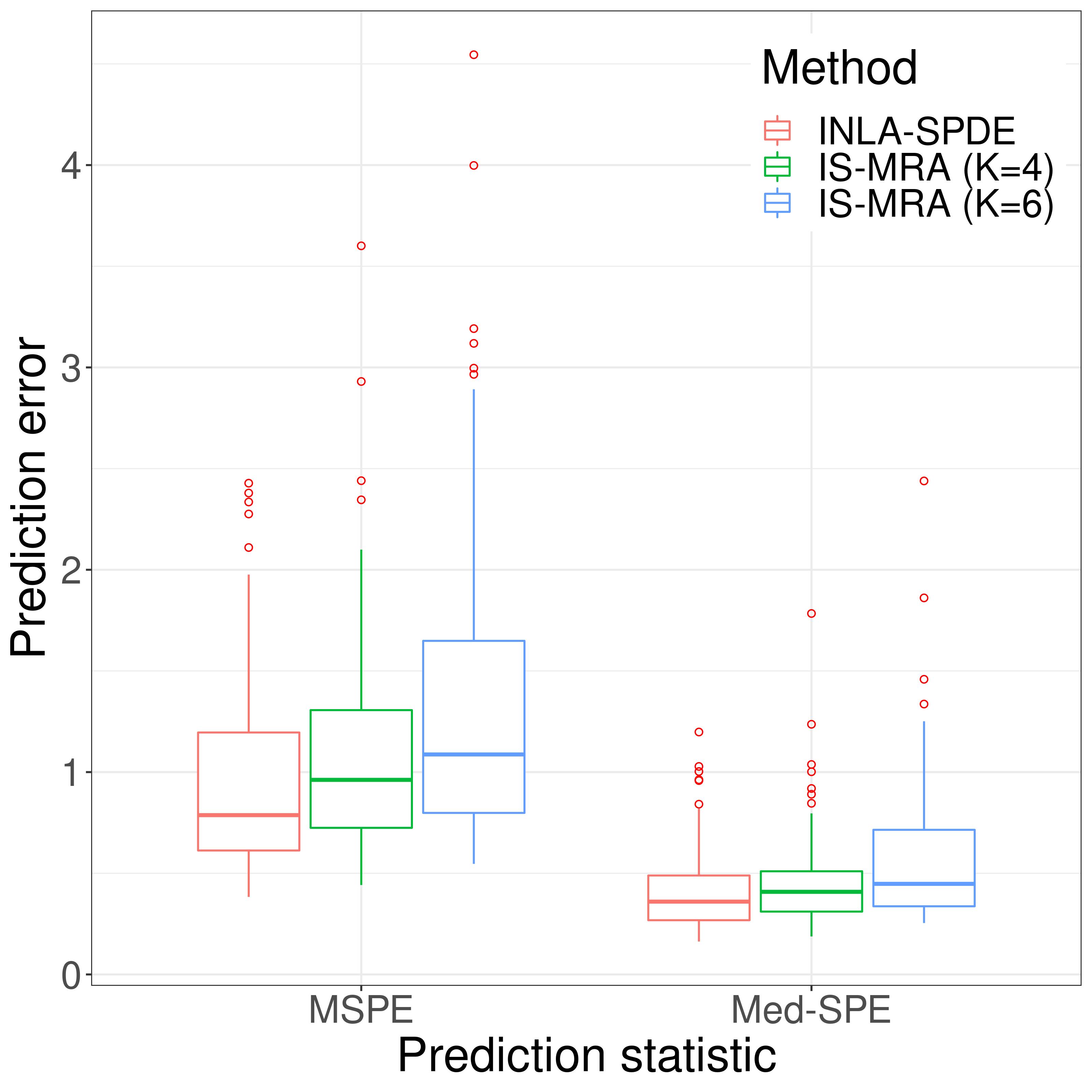}
    \caption{\textbf{Box plots for the mean squared prediction errors (MSPE) and the median squared prediction errors (Med-SPE) across simulated datasets.} Shown are results from INLA-SPDE (red) and IS-MRA with number of nested resolutions 4 (green) and 6 (blue).} 
    \label{fig:boxPlots}
\end{figure}

\begin{table}[ht]
\centering
\begin{tabular}{lccccc}
    \toprule
    & \multicolumn{1}{c}{\textbf{Mean}} & \multicolumn{1}{c}{\textbf{SD}} &
    \multicolumn{1}{c}{\textbf{Median}} &
    \multicolumn{1}{c}{\textbf{Min.}} & \multicolumn{1}{c}{\textbf{Max.}} \\ 
    \midrule
    \multicolumn{6}{l}{\textbf{IS-MRA (K=6)}} \\ 
    ~~MSPE & 1.3676 & 0.8027 & 1.0873 & 0.5466 & 4.5455 \\ 
    ~~Med-SPE & 0.5763 & 0.3657 & 0.4478 & 0.2542 & 2.4390 \\ 
    ~~Mean SD & 0.7702 & 0.0229 & 0.7704 & 0.7190 & 0.8419 \\ 
    ~~Med. SD & 0.7659 & 0.0223 & 0.7655 & 0.7170 & 0.8294 \\
    ~~$95$\% coverage prob. & 0.8342 & 0.1089 & 0.8700 & 0.4689 & 0.9615 \\
    \multicolumn{6}{l}{\textbf{IS-MRA (K=4)}} \\ 
    ~~MSPE & 1.1108 & 0.5539 & 0.9617 & 0.4420 & 3.6010 \\ 
    ~~Med-SPE & 0.4635 & 0.2401 & 0.4084 & 0.1877 & 1.7839 \\ 
    ~~Mean SD & 0.8412 & 0.0294 & 0.8408 & 0.6350 & 0.8968 \\ 
    ~~Med. SD & 0.8364 & 0.0302 & 0.8374 & 0.6126 & 0.8909 \\ 
    ~~$95$ \% coverage prob. & 0.9022 & 0.0716 & 0.9185 & 0.6264 & 0.9853 \\ 
    \midrule
    \multicolumn{6}{l}{\textbf{INLA-SPDE}} \\ 
    ~~MSPE & 0.9740 & 0.4840 & 0.7873 & 0.3830 & 2.4279 \\ 
    ~~Med-SPE & 0.4138 & 0.2053 & 0.3600 & 0.1628 & 1.1982 \\ 
    ~~Mean SD & 1.0657 & 0.0429 & 1.0685 & 0.9590 & 1.1742 \\ 
    ~~Med. SD & 1.0810 & 0.0432 & 1.0855 & 0.9746 & 1.1972 \\
     ~~$95$\% coverage prob. & 0.9542 & 0.0242 & 0.9615 & 0.8718 & 0.9872 \\
  \bottomrule 
\end{tabular}
\caption{\textbf{Summary of prediction statistics based on $100$ simulated datasets.} The $95$\% coverage probability corresponds to the proportion of $95$\% credible intervals for predicted values that encompass the true value.}  \label{tab:simPredStatistics}
\end{table}

Figure \ref{fig:boxPlots} and Table \ref{tab:simPredStatistics} let us compare predictions produced by IS-MRA and INLA-SPDE. MSPEs were moderately higher for IS-MRA than for INLA-SPDE. Reducing $K$ generally improved MSPE and Med-SPE. The comparatively smaller values in Med-SPE indicate that higher values tend to result from a small number of bad predictions, often from locations at which observed values include a large contribution from uncorrelated variation. INLA-SPDE and IS-MRA also differed in terms of the mean standard deviation across locations for the marginal posterior predictive distributions, with IS-MRA producing an average of $0.77$ at $K=6$ and $0.84$ at $K=4$, in comparison to $1.06$ for INLA-SPDE. IS-MRA underestimated the posterior predictive standard deviations, as mean coverage probabilities for predictions are $0.83$ and $0.90$ at $K=6$ and $K=4$, respectively. INLA-SPDE on the other hand, as expected, has mean coverage probability close to $95$\%. 

\begin{table}[ht]
\centering
\begin{tabular}{lrrr}
  \toprule
 & \textbf{SPDE} & \multicolumn{2}{c}{\textbf{IS-MRA}} \\
 & & $\bm{K = 6}$ & $\bm{K = 4}$ \\
  \midrule
  \multicolumn{4}{l}{\textbf{Fixed effects}} \\
  ~~Elevation & 0.96 & 0.96 & 0.96 \\ 
  ~~May 28 & 0.98 & 0.35 & 0.60 \\ 
  ~~May 29 & 0.93 & 0.25 & 0.48 \\ 
  ~~Evergreen broadleaf & 0.93 & 0.96 & 0.95 \\ 
  ~~Mixed forest & 0.92 & 0.96 & 0.95 \\ 
  ~~Closed shrublands & 0.94 & 0.94 & 0.94 \\ 
  ~~Savannas & 0.93 & 0.92 & 0.94 \\ 
  ~~Grasslands & 0.94 & 0.96 & 0.97 \\ 
  ~~Permanent wetlands & 0.95 & 0.98 & 0.96 \\ 
  ~~Croplands & 0.94 & 0.95 & 0.93 \\ 
  ~~Urban & 0.93 & 0.96 & 0.96 \\ 
  ~~Cropland-natural mosaics & 0.95 & 0.96 & 0.96 \\ 
  ~~Non-vegetated & 0.97 & 0.97 & 0.98 \\
  \multicolumn{4}{l}{\textbf{Hyperparameters}} \\
  ~~Spatial range ($\rho$) & 0.68 & 0.45 & 0.65 \\ 
  ~~Temporal range ($\phi$) & 0.00 & 0.65 & 0.81 \\ 
  ~~Std. dev. ($\sigma$) & 0.61 & 0.34 & 0.60 \\ 
  \bottomrule
\end{tabular}
\caption{\textbf{Coverage probabilities of 95\% credible intervals for fixed effects parameters and hyperparameters.}}
\label{tab:FEcoverageProbs}
\end{table}

\begin{figure}[p] 
    \begin{subfigure}[b]{0.48\textwidth}
    \centering
    \includegraphics[width = 5cm]{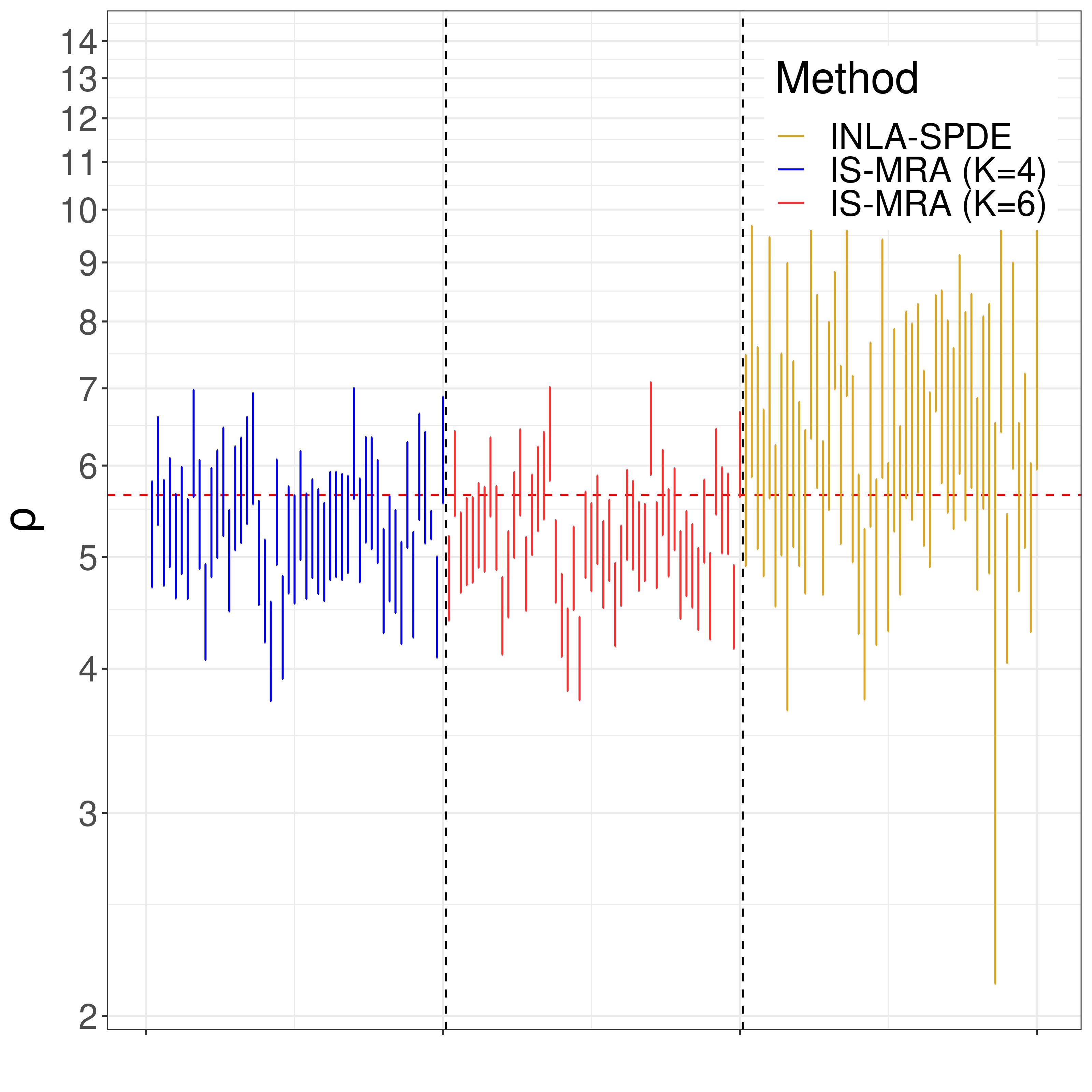}
    \caption{Spatial range} \label{fig:spaceRangeCredInts}
    \end{subfigure}
    \begin{subfigure}[b]{0.48\textwidth}
    \centering
    \includegraphics[width = 5cm]{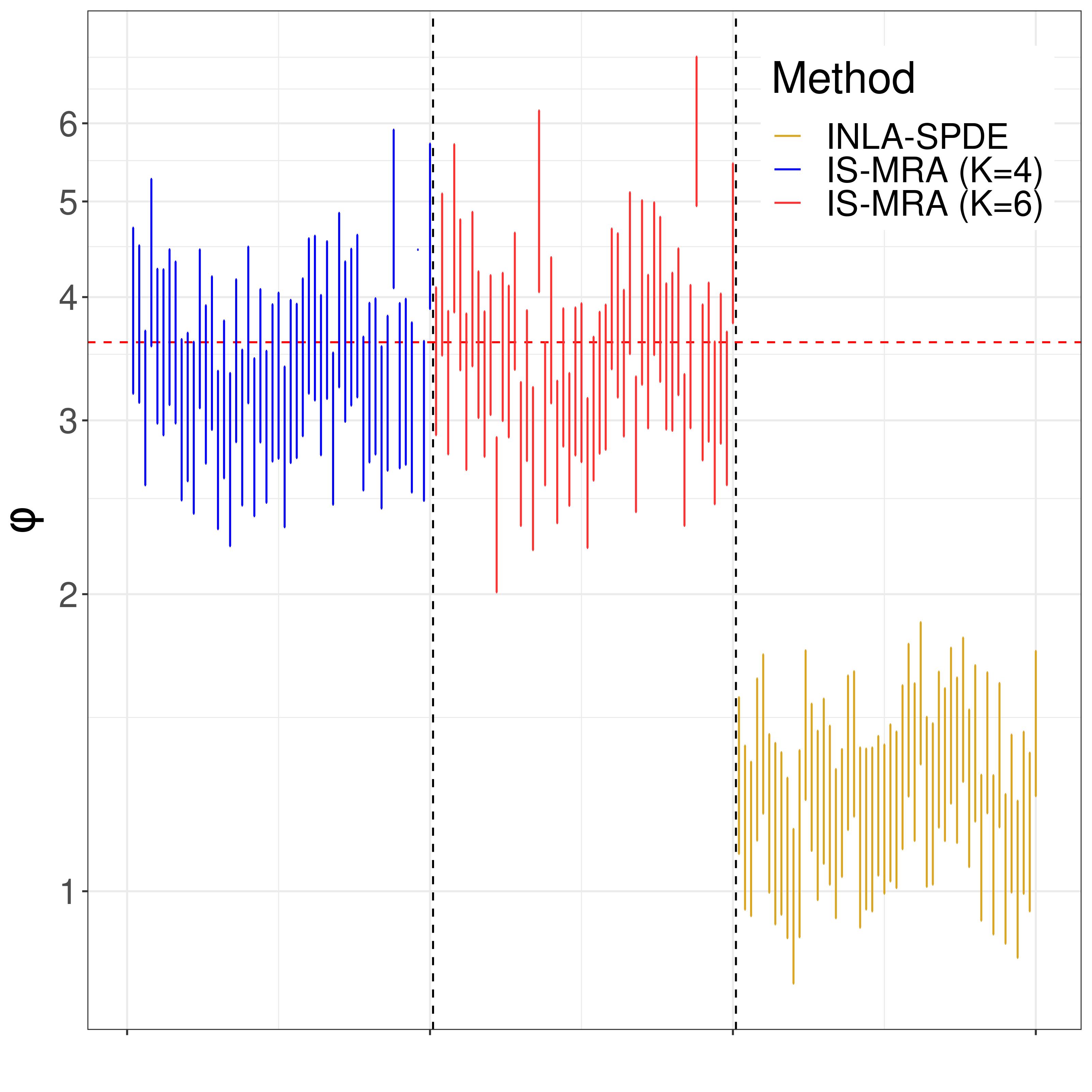}
    \caption{Temporal range} \label{fig:timeRangeCredInts}
    \end{subfigure}
    \begin{subfigure}[b]{0.48\textwidth}
    \centering
    \includegraphics[width = 5cm]{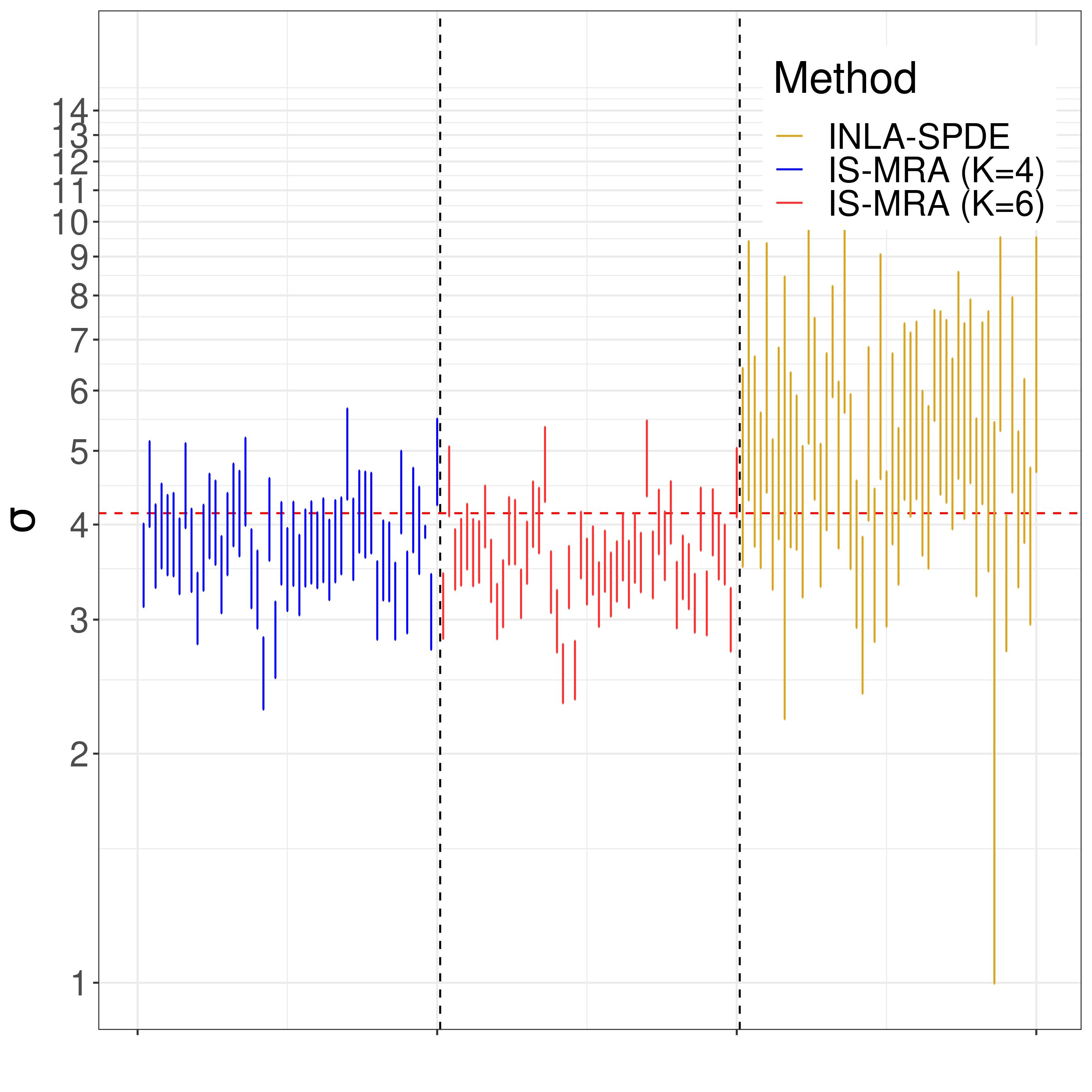}
    \caption{Std. dev.} \label{fig:scaleCredInts}
    \end{subfigure}
    \begin{subfigure}[b]{0.48\textwidth}
    \centering
    \includegraphics[width = 5cm]{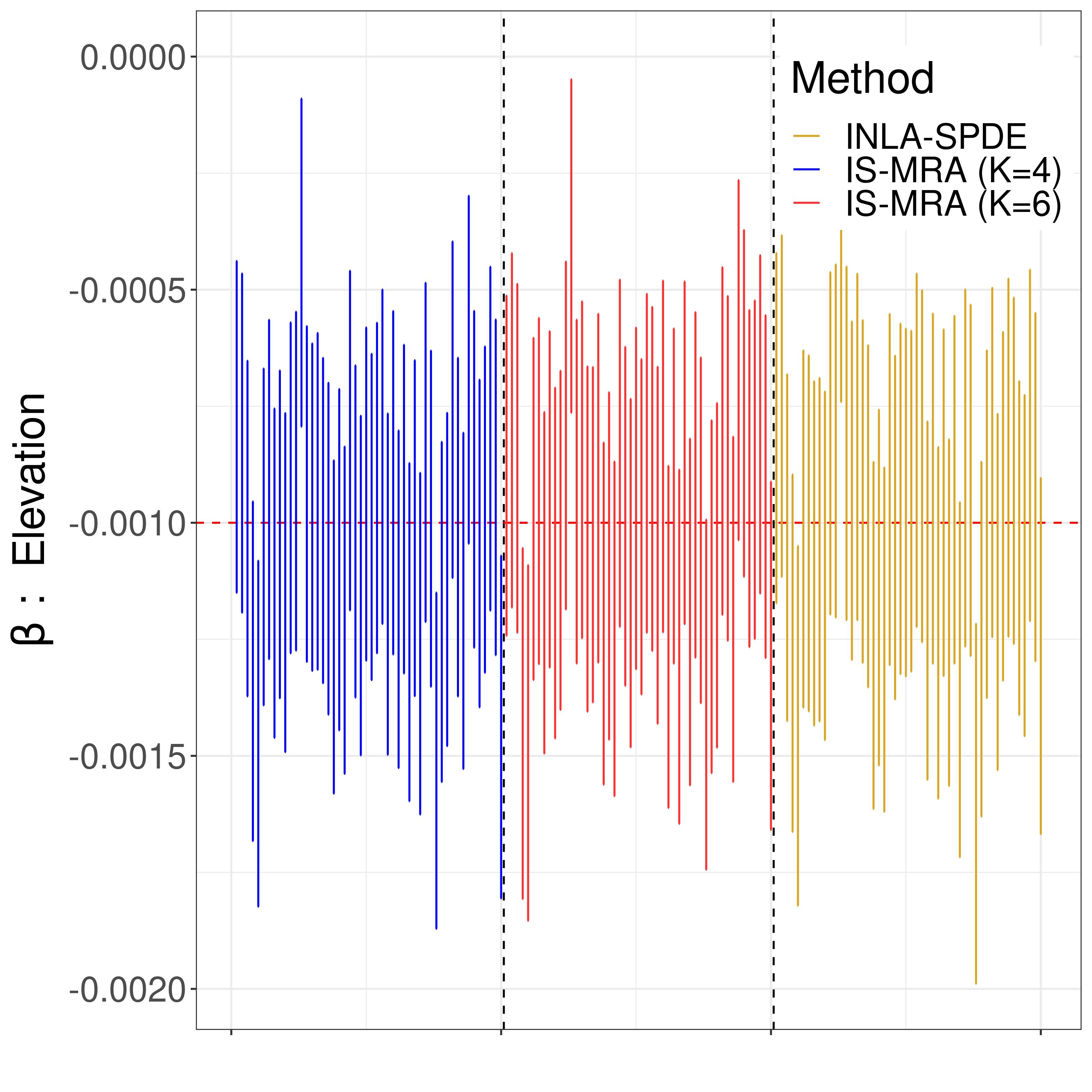}
    \caption{Fixed effect: Elevation} \label{fig:FEelevation}
    \end{subfigure}
    \begin{subfigure}[b]{0.48\textwidth}
    \centering
    \includegraphics[width = 5cm]{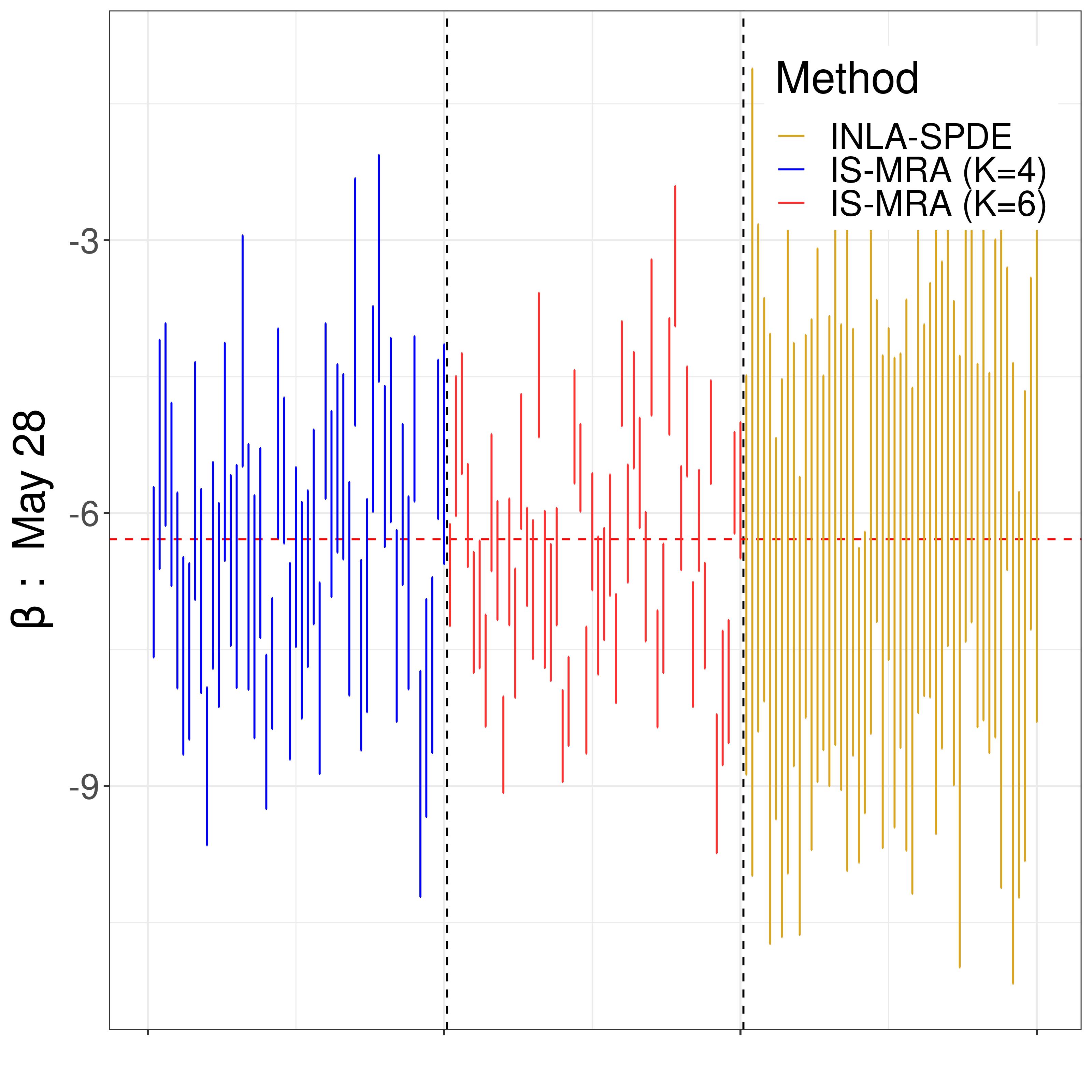}
    \caption{Fixed effect: May 28} \label{fig:FEmay28}
    \end{subfigure}
    \begin{subfigure}[b]{0.48\textwidth}
    \centering
    \includegraphics[width = 5cm]{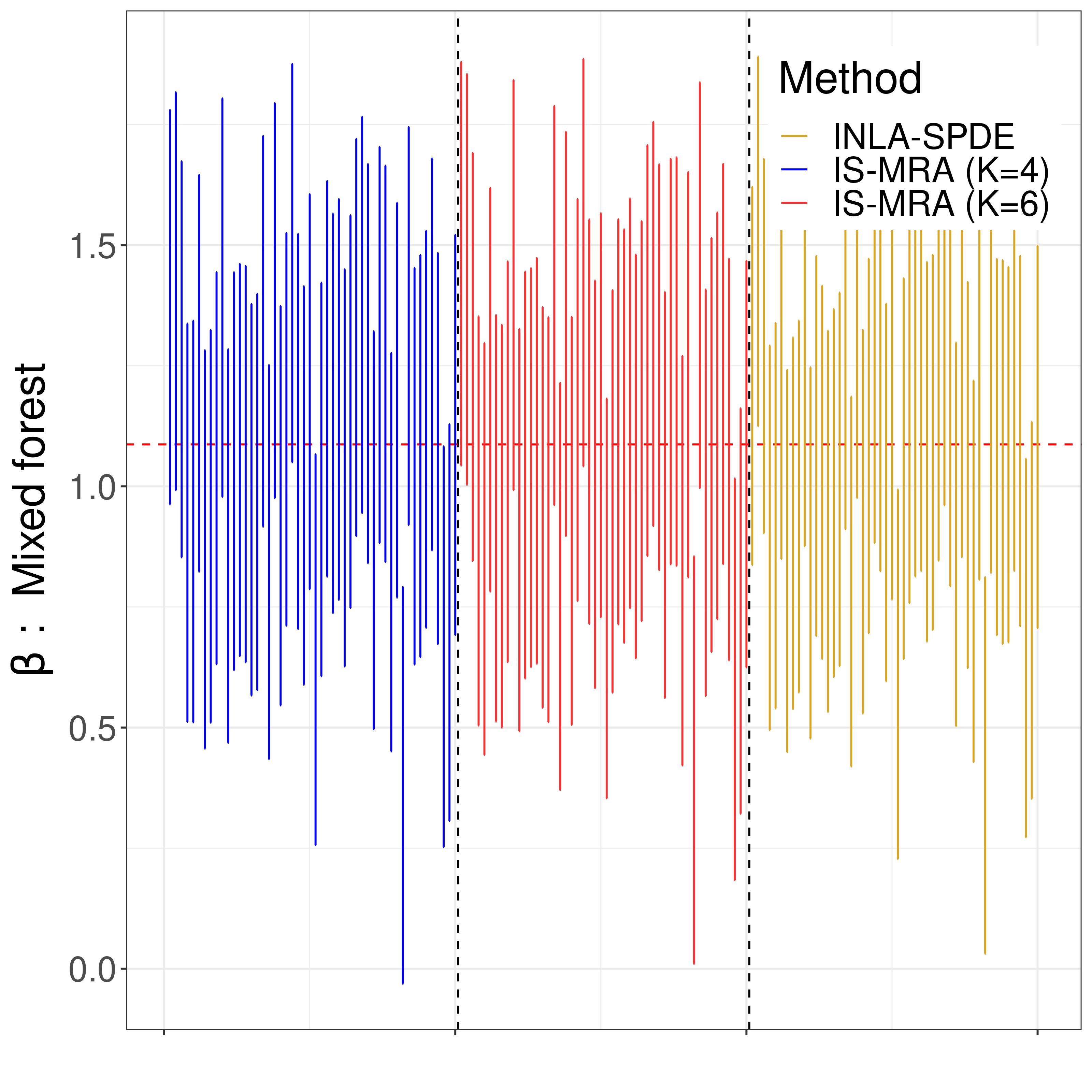}
    \caption{Fixed effect: Mixed forest} \label{fig:FEmixedForest}
    \end{subfigure}
    \caption{\textbf{95\% credible intervals for the hyperparameters and fixed effects.} Each bar represents an interval under a certain combination of method and simulated dataset. The horizontal red dashed line corresponds to the parameter's real value. We only show results for $50$ of the $100$ simulated datasets. The $y$ axis for graphs a), b), and c) is on the logarithmic scale. Bars are grouped in three bins, corresponding to the algorithm used and delimited by the vertical black lines. Their ordering with respect to the simulated datasets is the same in all three bins. It follows that, for example, the first blue, red, and yellow bars were all obtained from simulated dataset 1.}
    \label{fig:paraCredInts}
\end{figure}

IS-MRA also had a tendency to underestimate the standard deviation of posteriors for fixed effects and hyperparameters, as evidenced by Table \ref{tab:FEcoverageProbs}. Coverage probabilities were generally good for fixed effects parameters, except for the day of observation. Once again, we note an improvement when $K$ is smaller. INLA-SPDE and IS-MRA (K=4) both produced coverage probabilities of approximately $0.65$ and $0.60$ for the spatial range and standard deviation hyperparameters, respectively. INLA-SPDE had issues with the estimation of the temporal range hyperparameter though, with associated $95$\% credible intervals never including the real parameter value. IS-MRA ($K =4$), with coverage probability of $0.81$,  did much better.

The coverage probabilities we measured can be better understood by looking at Figure \ref{fig:paraCredInts}. Each bar in the graphs represents a credible interval obtained for a certain simulated dataset for a given method. For "elevation" and land cover effects such as "mixed forest", the credible intervals obtained with different methods were similar. For the "May 28" effect, INLA-SPDE produced much wider credible intervals than IS-MRA, which resulted in a much better coverage probability. A smaller $K$ once again resulted in wider credible intervals, and it was also the case for hyperparameters. The figures also highlight how INLA-SPDE struggles to estimate the temporal range hyperparameter, with credible intervals falling systematically below the real value. The small width of credible intervals for the temporal range result from the parameterisation assumed in INLA-SPDE, cf. Eq. \ref{eq:INLAtimeRange}. The simulated datasets include measurements on only three days, which we suspect explains the systematic underestimation.

The mean absolute differences between the real land cover effects values and corresponding posterior means were comparable for both algorithms, as can be seen in Table 1 in Appendix C. INLA-SPDE did better on average with the estimation of the time effects, but IS-MRA outperformed INLA-SPDE with the estimation of hyperparameters. Using a smaller $K$ did not systematically reduce the difference between posterior means and the real parameter values.

\section{Discussion}

The algorithm we devised, IS-MRA, allows computationally tractable fully Bayesian inference for large spatiotemporal datasets. The main improvement on existing MRA-based methods is enabling the computation of posterior distributions for covariance parameters through a numerical integration algorithm using importance sampling. The analyses demonstrate that IS-MRA can produce realistic and accurate land surface temperature predictions for a moderately large dataset collected in India.

Currently, the memory footprint of the algorithms used for obtaining the Cholesky decomposition of sparse matrices remains the greatest computational hurdle. Fortunately, under the MRA, the sparsity pattern of precision matrix $\bm{Q}$ does not depend on hyperparameters. Analysing and storing the sparsity pattern therefore results in important computational benefits, as it will be invariant across iterations of the IS algorithm. A strategy that leverages the sparsity structure of the different precision matrices could help resolve that issue, and help scale IS-MRA to datasets comprising tens of millions of observations. An extension to non-Gaussian likelihoods, point processes or categorical outcomes for example, would also be a welcome improvement. That extension would surely rely on the Integrated Nested Laplace Approximation (INLA) method \citep{Rue2009}. Further, we would need to refine the strategies for knot placement and partitioning the study region. Altering the MRA smoothing scheme for eliminating breaks in prediction surfaces would also be desirable. Reformulating our algorithm to use the general Vecchia approximation \citep{Katzfuss2020} instead of MRA would also be feasible, and might solve issues with edge effects.

A more thorough investigation of the theoretical computational performance of IS-MRA in comparison to INLA-SPDE was beyond the scope of this manuscript, but would be helpful. Unfortunately, comparing both approaches systematically on that front is challenging, given the number of tuning parameters involved in the fitting of each underlying model and their potential effects on predictive accuracy. For example, allowing more CPUs to be used for INLA-SPDE estimation might greatly increase the computational speed, while simultaneously requiring more memory. Alternatively, a surefire way of speeding up IS-MRA would be to reduce the number of knots, but doing so may lessen predictive accuracy. Moreover, unlike IS-MRA, INLA-SPDE estimation relies on software with a long development history. In other words, it has benefited from extensive profiling work, unlike the in-house solution provided with this paper, putting it at a clear disadvantage.

Massive datasets remain a considerable challenge for Bayesian inference methods. Nevertheless, their capacity to intuitively quantify uncertainty in parameter estimates or predictions and to account for measurement error in observations can be very valuable in practice. IS-MRA represents a worthy step towards scaling Bayesian inference to much larger spatiotemporal datasets. The proposed improvements will help make the algorithm more flexible, and ultimately, applicable to data of a scale comparable to that of the MODIS land surface temperature database.   

\bibliographystyle{ba}
\bibliography{postdocBiblio}

\section*{Acknowledgements}

The authors would like to thank Nancy Reid for her invaluable help in reviewing the manuscript. This work was funded by the Canadian Statistical Sciences Institute (CANSSI); the Institut de Valorisation des Données (IVADO) under Grant PRF-2017-02; and NSERC under Grants RGPIN-2017-06856 and RGPIN-2016-06296.

\newpage

\appendix
\renewcommand\thefigure{\thesection.\arabic{figure}}
\setcounter{figure}{0}
\setcounter{table}{0}

\section{Additional notes on the implementation of IS-MRA}

The importance sampling strategy we described involves running a short optimisation procedure to estimate the mode of $p[\bm{\Psi} \mid \bm{y}]$. The software completes that step with the low-memory BFGS algorithm \citep{Liu1989}. Since there is no closed-form expression for the gradient, we rely on numerical differentiation instead. From the user-provided starting values, the software performs by default $25$ iterations. 

The software allows the user to specify the number of longitude, latitude, and time splits required. Longitude splits are processed first, followed by latitude splits, and finally, time splits. For example, if we require one longitude, one latitude, and one time split, we'll have $K = 3$. The algorithm first splits the entire spatiotemporal domain in two, based on observation longitudes, creating resolution $1$. Then, the algorithm splits each of the resulting two subregions in two, based on observation latitudes, creating resolution $2$. Finally, it splits all subregions in resolution $2$ based on the observations' temporal coordinates, resulting in resolution 3. 

To ensure a good balance in the distribution of observations among subregions, the software places the new boundary for each split at the median value for the dimension to be split. Boundaries are left-continuous, that is, an observation on a boundary is assigned to the subregion on its left ($\leq$). The total number of knots at each resolution grows by a multiplicative constant. By default, the software sets the number of knots at $20$ at resolution $0$, and multiplies it by a factor of $J = 2$ at each resolution until $M - 1$. In practice, more knots may reduce MSPE, but especially at high resolutions, adding knots tends to sizably increase the computational burden. That is why we also included a tuning parameter called the \textit{tip knots thinning rate}, comprised between $0$ and $1$. If we set this tuning parameter at $0.5$, for example, the software will only retain $50$\% of knots in each subregion at the finest resolution, the selection of which is uniform at random. By selecting a suitable thinning rate, we can keep the full conditional precision matrix $\bm{Q}$ at a size that can be handled by the LDLT decomposition algorithm offered in the Eigen library. We stress that configuring the algorithm to respect memory constraints requires mainly limiting the number of knots, and the thinning rate is an important feature for that purpose.

Under certain hyperparameter values, the covariance matrices computed for the $\bm{\eta}$ parameter vectors in the MRA can be computationally singular. To prevent the issue, we apply by default a nugget effect of $1e-5$ to both the spatial and temporal covariance functions. \newpage

\section{Notes on the application} 

The priors for all fixed effects $\bm{\beta}$ are normal with mean $\bm{0}$ and covariance $100 \bm{I}$, a large variance with respect to the scale of the fixed effects considered. All hyperparameters are expressed on the logarithmic scale, and have normal hyperpriors with mean $0$ and standard deviation $2$. We deemed that such a standard deviation would allow for a suitable range of probable values, since $[\exp(-4), \exp(4)] \approx [0.02, 54.60]$. We are very confident those bounds encompass the range or scaling values one would expect in LST data. In other words, those values were selected arbitrarily to make the priors reasonably uninformative. The standard deviation for the measurement error term, $\log(\sigma_{\epsilon})$, known from validation studies \citep{Wan2014}, is fixed at $\log(0.5)$. We therefore have only three variable hyperparameters: the spatial and temporal range log-parameters, $\log(\rho)$ and $\log(\phi)$, and the standard deviation log-parameter $\log(\sigma)$. We center all continuous covariates.

In the main analysis, we create six longitude splits, five latitude splits, and one time split, which results in $M = 12$. We place $8$ knots in each region from resolutions $0$ to $11$ ($= M-1$) based on the default placement scheme. We sample $100$ values in the importance sampling step. We let the optimizer, used to identify the mode of the joint marginal hyperparameters posterior distribution, run for $20$ iterations. We impose a tip knots thinning rate of one third. All those tuning parameters control the computational burden of the algorithm. The total number of splits, $M$, should be set as small as possible, keeping in mind however that a smaller $M$ results in larger matrices to invert in the computation of the $\bm{\Gamma}.$ matrices, and affects the sparsity of $\bm{H}(\bm{\Psi}_{ST}; \bm{\mathcal{Q}})$. Setting $M = 12$ ensured that the algorithm would run fairly quickly, and not use more than the available memory. Setting it lower would have greatly increased running time. Since we only had seven days of data, we deemed that we should not have more than one time split. We divided the remaining splits, $11$, in two, resulting in the six longitude and five latitude splits mentioned. Because the default knot placement scheme starts by placing knots on the eight vertices of a rectangular prism, we thought that $8$ would be the minimum number of knots recommended. That choice ensures that prediction locations are always close to the selected knots at any resolution, and results in a smaller $\bm{H}(\bm{\Psi}_{ST}; \bm{\mathcal{Q}})$ matrix. Finally, the thinning rate of one third eliminated $733,334$ columns in $\bm{H}(\bm{\Psi}_{ST}; \bm{\mathcal{Q}})$. With that choice, we made sure that the algorithm would not request more memory than was available. The re-ordering scheme implemented in the SimplicialLDLT solver in Eigen (cf. \textit{analyzePattern}) is especially memory-intensive, and so, we found it best to keep $\bm{H}(\bm{\Psi}_{ST}; \bm{\mathcal{Q}})$ below $700,000$ columns.

In the validation analysis, as the region surveyed is smaller, we consider instead four longitude splits, five latitude splits, and no time split, which results in $M = 9$. We apply a tip knots thinning rate of $0.5$, and all other tuning parameters are the same. Once again, those tuning parameters were selected for computational reasons: we found that they offered a reasonable trade-off between computational and predictive performance.

\newpage

\section{Additional graphs and tables} 

\begin{figure}[h]
    \centering
    \includegraphics[width = 13cm]{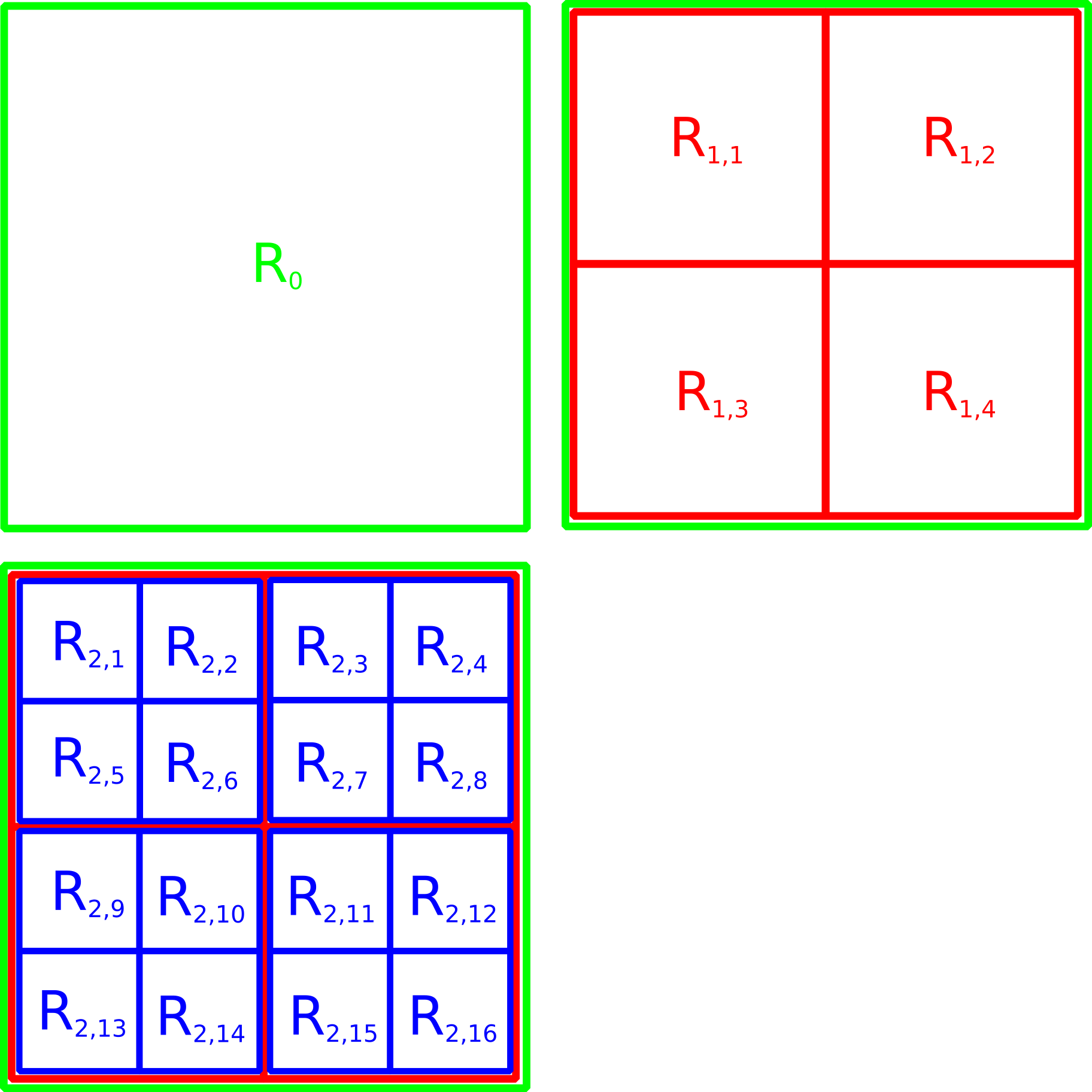}
    \caption{\textbf{Example of the nesting structure for the MRA.} $K$ is equal to three, and each resolution has a different colour. The nesting in grids is not reflected in the $R_{x,y}$ notation.}
    \label{fig:resolution}
\end{figure}

\begin{figure}[p]
    \centering
    \includegraphics[width = 13cm]{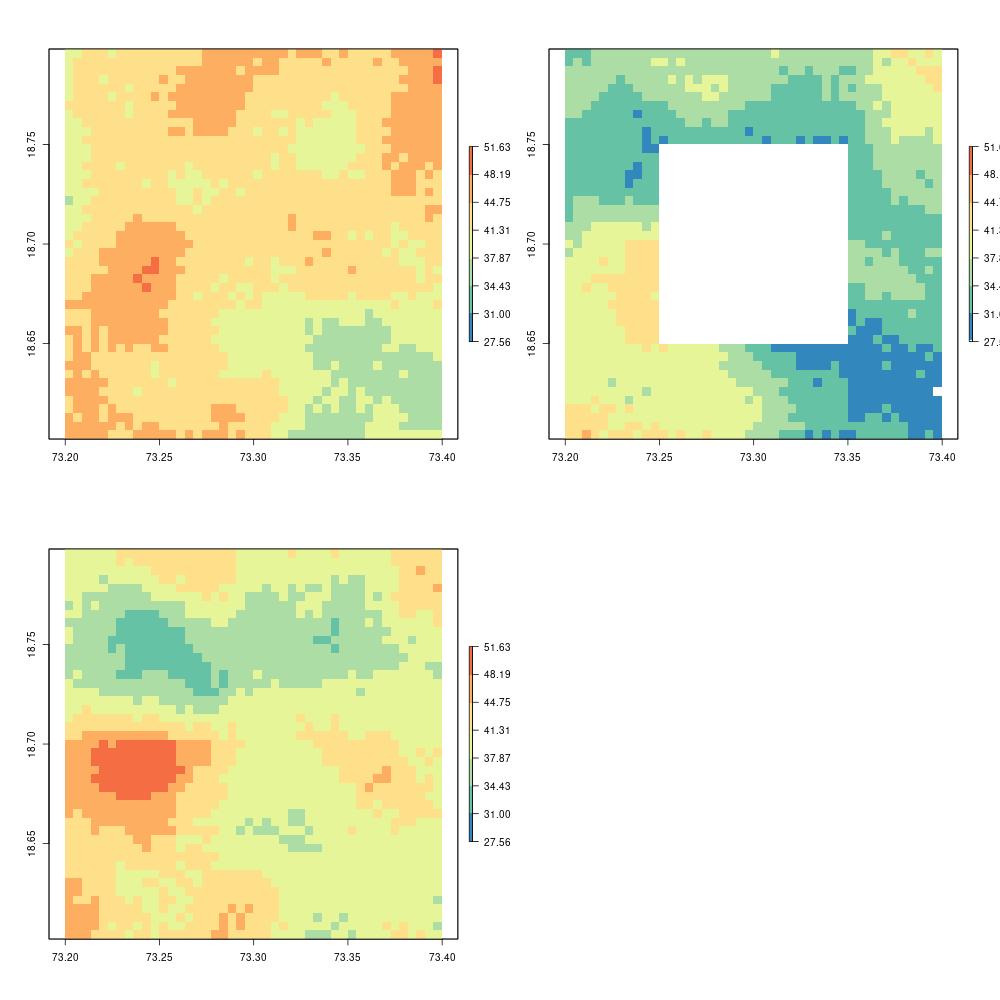}
    \caption{\textbf{Training dataset used in the simulation study.} Each map is for a different day. The white block in the middle of the second map represents the data subtracted for validation.}
    \label{fig:simDatasets}
\end{figure}

\begin{table}[ht]
    \begin{center}
    \resizebox{0.95\textwidth}{!}{
    \begin{tabular}{lrrrrrrrrr}
    \toprule
    & \multicolumn{3}{c}{\textbf{Mean}} & \multicolumn{3}{c}{\textbf{Min.}} & \multicolumn{3}{c}{\textbf{Max.}} \\
    & \multicolumn{2}{c}{\textbf{IS-MRA}} & \textbf{SPDE} & \multicolumn{2}{c}{\textbf{IS-MRA}} & \textbf{SPDE} & \multicolumn{2}{c}{\textbf{IS-MRA}} & \textbf{SPDE} \\
    & \textbf{K=4} & \textbf{K=6} &  & \textbf{K=4} & \textbf{K=6} &  & \textbf{K=4} & \textbf{K=6} & \\
    \midrule
    \multicolumn{10}{l}{\textbf{Fixed effects}} \\
    ~~Elevation & 0.0001 & 0.0002 & 0.0002 & <0.0001 & <0.0001 & <0.0001 & 0.0006 & 0.0006 & 0.0006 \\ 
    ~~May 28 & 1.0195 & 1.0219 & 0.8358 & 0.0418 & 0.0450 & 0.0446 & 3.0165 & 3.0992 & 2.1954 \\ 
    ~~May 29 & 1.5453 & 1.6334 & 1.3191 & 0.0173 & 0.0081 & 0.0283 & 6.0215 & 6.2216 & 5.2348 \\ 
    ~~Evergreen broadleaf & 0.1999 & 0.2132 & 0.1983 & 0.0022 & 0.0017 & 0.0043 & 0.6850 & 0.7837 & 0.6378 \\ 
    ~~Mixed forest & 0.1717 & 0.1720 & 0.1660 & 0.0088 & 0.0012 & 0.0029 & 0.7067 & 0.6543 & 0.6652 \\ 
    ~~Closed shrublands & 0.1859 & 0.1834 & 0.1816 & 0.0056 & 0.0005 & 0.0008 & 0.6379 & 0.5821 & 0.6524 \\ 
    ~~Savannas & 0.1991 & 0.2135 & 0.1950 & <0.0001 & 0.0005 & 0.0024 & 0.6758 & 0.6250 & 0.6910 \\ 
    ~~Grasslands & 0.1737 & 0.1700 & 0.1730 & 0.0013 & 0.0013 & 0.0122 & 0.5478 & 0.5514 & 0.5470 \\ 
    ~~Permanent wetlands & 0.1656 & 0.1631 & 0.1625 & 0.0139 & 0.0007 & 0.0019 & 0.5280 & 0.5685 & 0.5151 \\ 
    ~~Croplands & 0.2649 & 0.2648 & 0.2549 & 0.0075 & 0.0060 & 0.0026 & 0.9148 & 0.8904 & 0.8490 \\ 
    ~~Urban & 0.1638 & 0.1617 & 0.1612 & 0.0042 & 0.0053 & 0.0007 & 0.5407 & 0.5268 & 0.5401 \\ 
    ~~Cropland-natural mosaics & 0.1665 & 0.1606 & 0.1656 & 0.0066 & <0.0001 & 0.0003 & 0.5062 & 0.5108 & 0.5110 \\ 
    ~~Non-vegetated & 0.1759 & 0.1777 & 0.1717 & 0.0126 & 0.0065 & 0.0001 & 0.6773 & 0.7018 & 0.6929 \\ \midrule
    \multicolumn{10}{l}{\textbf{Hyperparameters}} \\
    ~~Spatial range ($\rho$) & 0.1169 & 0.1062 & 0.1490 & <0.0001 & 0.0026 & 0.0011 & 2.8970 & 0.3254 & 0.5433 \\ 
    ~~Temporal range ($\phi$) & 0.1030 & 0.1512 & 1.6920 & 0.0021 & 0.0002 & 1.5631 & 0.3365 & 0.5009 & 1.8370 \\ 
    ~~Std. dev. ($\sigma$) & 0.1533 & 0.1473 & 0.8920 & 0.0005 & 0.0018 & 0.3455 & 3.3072 & 0.4813 & 1.5329 \\
   \bottomrule
\end{tabular}
}
\end{center}
\caption{\textbf{Summary of absolute difference between the mean of the parameter posterior estimate and real values across simulated datasets.} Hyperparameters are on the logarithmic scale.}
\label{tab:absErrors}
\end{table}

\end{document}